\documentclass[aps,prd,superscriptaddress,nofootinbib,amsmath,amsfonts,preprintnumbers,groupedaddress,10pt,english]{revtex4}
\usepackage{amsmath}
\usepackage{amssymb}
\usepackage{babel}
\usepackage{wrapfig}
\usepackage{cancel}

\usepackage{relsize,exscale}
\makeatletter

\usepackage{array,multirow,graphicx}
\usepackage{dcolumn}
\usepackage{newlfont}
\usepackage{bm}
\usepackage[colorlinks,citecolor=blue,urlcolor=blue,linkcolor=blue]{hyperref}
\usepackage[figtopcap]{subfigure}
\usepackage{color}
\usepackage{verbatim}

\usepackage{scalerel}
\usepackage{tikz}
\usetikzlibrary{svg.path}
\definecolor{orcidlogocol}{HTML}{A6CE39}
\tikzset{
  orcidlogo/.pic={
    \fill[orcidlogocol] svg{M256,128c0,70.7-57.3,128-128,128C57.3,256,0,198.7,0,128C0,57.3,57.3,0,128,0C198.7,0,256,57.3,256,128z};
    \fill[white] svg{M86.3,186.2H70.9V79.1h15.4v48.4V186.2z}
                 svg{M108.9,79.1h41.6c39.6,0,57,28.3,57,53.6c0,27.5-21.5,53.6-56.8,53.6h-41.8V79.1z M124.3,172.4h24.5c34.9,0,42.9-26.5,42.9-39.7c0-21.5-13.7-39.7-43.7-39.7h-23.7V172.4z}
                 svg{M88.7,56.8c0,5.5-4.5,10.1-10.1,10.1c-5.6,0-10.1-4.6-10.1-10.1c0-5.6,4.5-10.1,10.1-10.1C84.2,46.7,88.7,51.3,88.7,56.8z};}}
\newcommand\orcid[1]{\href{https://orcid.org/#1}{\mbox{\scalerel*{
\begin{tikzpicture}[yscale=-1,transform shape]
\pic{orcidlogo};
\end{tikzpicture}
}{|}}}}
\graphicspath{{./}{Figs/}}
\begin{document}
\date{\today}

\title{Constraining Quadratic {$f(R)$} Gravity from Astrophysical Observations of the Pulsar J0704+6620}

\author{G.~G.~L.~Nashed~\orcid{0000-0001-5544-1119}}
\author{Waleed El~Hanafy~\orcid{0000-0002-0097-6412}}
\email{nashed@bue.edu.eg}
\email{waleed.elhanafy@bue.edu.eg}
\affiliation {Centre for Theoretical Physics, The British University, P.O. Box
43, El Sherouk City, Cairo 11837, Egypt}



\begin{abstract}
We apply quadratic $f(R)=R+\epsilon R^2$ field equations, where $\epsilon$ has a dimension [L$^2$], to static spherical stellar model. We assume the interior configuration is determined by Krori-Barua ansatz and additionally the fluid is anisotropic. Using the astrophysical measurements of the pulsar PSR J0740+6620 as inferred by NICER and XMM observations, we determine $\epsilon\approx \pm 3$ km$^2$. We show that the model can provide a stable configuration of the pulsar PSR J0740+6620 in both geometrical and physical sectors.  We show that the Krori-Barua ansatz within $f(R)$ quadratic gravity provides semi-analytical relations between radial, $p_r$, and tangential, $p_t$, pressures and density $\rho$ which can be expressed as $p_r\approx v_r^2 (\rho-\rho_1)$ and $p_r\approx v_t^2 (\rho-\rho_2)$, where $v_r$ ($v_t$) is the sound speed in radial (tangential) direction, $\rho_1=\rho_s$ (surface density) and $\rho_2$ are completely determined in terms of the model parameters. These relations are in agreement with the best-fit equations of state as obtained in the present study. We further put the upper limit on the compactness, which satisfies the $f(R)$ modified Buchdahl limit. Interestingly, the quadratic $f(R)$ gravity with negative $\epsilon$ naturally restricts the maximum compactness to values lower than Buchdahl limit, unlike the GR or $f(R)$ gravity with positive $\epsilon$ where the compactness can arbitrarily approach the black hole limit $C\to 1$. The model predicts a core density a few times the saturation nuclear density $\rho_{\text{nuc}} = 2.7\times 10^{14}$ g/cm$^3$, and a surface density $\rho_s > \rho_{\text{nuc}}$. We provide the mass-radius diagram corresponding to the obtained boundary density which has been shown to be in agreement with other observations.
\end{abstract}

\maketitle

\section{Introduction}\label{Sec:Introduction}

For many decades the detection of radio pulse times of arrival from pulsars is being used to measure their masses via Shapiro time delay. In particular, millisecond pulsars--- whose rotational frequency range 33--719 Hz, slow down rate $\leq 10^{-19}$ s/s and characteristic age up to Gyrs--- provide perfect laboratories to test relativistic gravity \cite{Stairs:2003eg,Reardon:2015kba}. However, the radii measurements are relatively difficult. Recently, Interior Composition Explorer (NICER) observations open a new window to measure pulsars radius by observing the X–ray light curves generated by rotating hot spots on the pulsars' surfaces in addition to bending of light \citep{Bogdanov:2019ixe,Bogdanov:2019qjb}. In addition, observations of gravitational wave signals by Laser Interferometer Gravitational-Wave Observatory (LIGO) and Virgo collaboration provide a new tool to measure pulsar radius \citep{Abbott:2016blz}.

It is widely believed that pulsars are neutron stars (NS) where the core matter consists of neutrons. These objects are characterized by high dense matter few times the nuclear saturation density $\rho_\text{nuc} \approx 2.7 \times 10^{14}$ g/cm$^3$, intense magnetic field $10^{12}$ Gauss, rapid rotation 1.39 ms. Others suggest that pulsars with high masses $\sim 2 M_\odot$ could have quark cores \cite{Bhattacharyya:2016kte,Annala:2019puf}. A pure quark star is possible, if strange quark matter is the true ground state of strongly interacting matter as conjectured by \citet{Witten:1984rs}, see  also \citep{Farhi:1984qu}. This implies a new minimum of the energy per baryon, at zero pressure, lower than the energy per baryon of Iron nuclei $^{56}$Fe ($ < 930~\text{MeV}$). On the other hand, pulsars provide strong gravitational field regimes where their masses and radii are estimated as $M\approx 1.5 M_\odot$ and $R\approx 10$ km. In this sense pulsars may represent the most exciting and unique observed objects in nature. Many studies which combine nuclear physics, solid state, stellar structure physics are devoted to study possible interactions inside these stars, their origin, formation, and evolution. Since the pulsar density limit is not accessible by terrestrial laboratories, the equation of state (EoS) of pulsars' matter is still unreachable \citep{Ozel:2016oaf}. One way to constrain the EoS inside the pulsar is to measure its mass and radius simultaneously.

The unprecedented progress in astrophysical observations of pulsars' masses and radii by combining Shapiro time delay (radio signals), X-rays and gravitational wave signals provides a powerful constraint on the proposed EoSs. We mention those mass-radius measurements: The pulsar PSR J0740+6620 with mass $M= 2.08 \pm 0.07 M_\odot$ \citep{NANOGrav:2019jur,Fonseca:2021wxt} and radius $R= 13.7_{-1.5}^{+2.6}$ km \cite{Miller:2021qha} another independent NICER measurement $R=12.39_{-0.98}^{+1.30}$ km \cite{Riley:2021pdl}, the PSR J0348+0432 with mass $M= 2.01 \pm 0.04 M_\odot$ and an estimated radius $R=13 \pm 2$ km \citep{Antoniadis:2013pzd} and the PSR J1614\textendash{2230} with mass $M= 1.908 \pm 0.016 M_\odot$ and radius $R=13 \pm 2$ km \citep{Demorest:2010bx,Fonseca:2016tux,NANOGRAV:2018hou}. Also, the PSR J0030+0451 with mass $M=1.44^{+0.15}_{-0.14} M_\odot$ and radius $R= 13.02_{-1.06}^{+1.24}$ km as measured by NICER \citep{Miller:2019cac}, with another independent NICER measurement with mass $M= 1.34^{+0.15}_{-0.16} M_\odot$ and radius $R= 12.71^{+1.14}_{-1.19}$ km \citep{Raaijmakers:2019qny}. The PSR J0437\textendash{4715} with mass $M=1.44 \pm 0.07 M_\odot$ \citep{Reardon:2015kba} and radius $R=13.6 \pm 0.9$ km \citep{Gonzalez-Caniulef:2019wzi} by the analyses of the surface X-ray thermal emission. Moreover, we include three observed mass\textendash{radius} values as inferred by gravitational wave signals as detected by LIGO/Virgo collaboration: The first detected NS-NS merger GW170817-1 with mass $M=1.45 \pm 0.09 M_\odot$ and radius $R=11.9 \pm 1.4$ km, and GW170817-2 with mass $M=1.27 \pm 0.09 M_\odot$ and radius $R=11.9 \pm 1.4$ km \citep{LIGOScientific:2018cki}. The LIGO/Virgo constraints on the radius of a canonical NS using GW170817+GW190814 signals with mass $M=1.4 M_\odot$ and radius $R=12.9\pm 0.8$ km \citep{LIGOScientific:2020zkf}. Furthermore, we include the mass and radius measurements of the possibly lightest neutron star, $M=0.77_{-0.17}^{+0.20}~M_{\odot}$ and $R=10.4^{+0.86}_{-0.78}$ km, located within the supernova remnant HESS J1731\textendash{347} based on the X-ray spectrum and distance estimate as obtained by Gaia observations \citep{2022NatAs...6.1444D}. On the other extreme the PSR J0952\textendash{0607}, its mass $M = 2.35\pm 0.17\, M_\odot$ represents the possibly heaviest neutron star ever observed \cite{Romani:2022jhd} with an estimated radius $R=14.087 \pm 1.0186$~km \cite{2023arXiv230514953E}.

Notably, the stellar evolution of isotropic models leave a mass gap between heaviest NS and lightest black hole (BH), $2.2 M_\odot \lesssim M \lesssim 5M_\odot$, unpopulated \citep{Yang:2020xyi}. The isotropic fluid assumption, radial  and tangential pressures are assumed to be equal ($p_r=p_t$), could be useful as an approximation. At high dense matter as in the pulsar core strong anisotropy is more realistic and naturally expected due to superfluidity, solidification, strong magnetic fields, hyperons, quarks as well as pion and kaon condensation. We also note that strong anistotropy induces an additional repulsive force inside the pulsar which allows for higher compactness value near the BH limit $C=2GM/c^2R \to 1$ when general relativity (GR) is considered \cite{Alho:2022bki}. In this case, additional physical constraints are needed to suppress this value to physical limit \cite{Alho:2021sli,Roupas:2020mvs,Raposo:2018rjn,Cardoso:2019rvt}. Similarly, this can be extended to modified gravity when matter--geometry nonminimal coupling is considered \citep{Nashed:2022zyi,ElHanafy:2022kjl,2023arXiv230514953E}.

From the weak field regime of solar system to the strong field of NSs, GR has passed several tests. At the same time, it is essential to test possible deviations by accounting for possible modifications on the gravitational sector. In this regard, the accurate observational data plays an extremely important role to constrain the parameter space of possible models. In particular, a well motivated and extensively studied extension of GR generalizes Einstein-Hilbert action to include an arbitrary function of Ricci scalar $f(R)$ instead of $R$. Several NS models have been constructed and investigated within $f(R)$ gravity \cite{Kobayashi:2008tq,Upadhye:2009kt,Feng:2017hje,TeppaPannia:2016vsb,Wojnar:2016bzk,Arapoglu:2016ozr,Katsuragawa:2015lbl,Fiziev:2015xpa,Hendi:2015pua,Momeni:2015vwa,Zubair:2016kov,Bakirova:2016ffk,AparicioResco:2016xcm,Moraes:2015uxq,Sharif:2015jaa,Sotani:2017pfj,Capozziello:2011nr,Arapoglu:2010rz,Astashenok:2013vza,Astashenok:2014pua,Astashenok:2014gda,Astashenok:2014nua}. It is the aim of the present study to constrain quadratic $f(R)=R+\epsilon R^2$ modified gravity from astrophysical observations of mass and radius of the pulsar J0740+6620 and to investigate the stability of the obtained solution. In this study, we use the more realistic case of strong anisotropic matter fields, i.e. $p_t> p_r$, to describe the matter inside the pulsar. We note that, as in the standard case, one needs to impose an extra condition by considering an EoS (two EoSs in anisotropic case) in order to solve the field equations. We instead apply the Krori-Barua (KB) ansatz \cite{Krori1975ASS} to the metric potentials in order to constrain the field equations.\\

The arrangement of this study is as follows: In Sec. \ref{Sec:fR_gravity}, we review the basic formalism of the $f(R)$ modified gravity and its field equations for a spherically symmetric spacetime configuration whereas the matter is assumed to be anisotropic. In Sec. \ref{Sec:Model}, we apply quadratic $f(R)=R+\epsilon R^2$ gravity, Starobinsky gravity, to KB stellar model, in addition to matching conditions with Schwarzschild exterior vacuum solution. In Sec. \ref{Sec:Stability}, we use the astrophysical observations of the mass and radius of the pulsar J0740+6620 to constrain the model parameter $\epsilon$. Moreover, we examine the validity of the model via several stability conditions on both geometry and matter sectors. In Sec. \ref{Sec:EoS_MR}, we obtain induced EoSs which govern the matter sector. We also discuss the modified Buchdahl limit on the compactness in the quadratic gravity theory in addition to the corresponding mass-radius diagram. In Sec. \ref{Sec:Conclusion}, we conclude the present study.
\section{Basic formalism of $f(R)$ gravity}\label{Sec:fR_gravity}

Let the pair ($\mathcal{M}, g$) be a Riemann manifold where $\mathcal{M}$ is a four-dimensional smooth manifold that admits a Riemannian metric $g$. One of the natural extensions to the GR theory can be realized by replacing the Ricci scalar, $R$, in Einstein-Hilbert action to be a general function $f(R)$. Therefore, the action of $f(R)$ modified gravity takes the form
\begin{equation}\label{action}
  {  {\cal S} = \int d^4x\sqrt{-g} \left[ \frac{1}{2\kappa}f(R) + {L}_m  \right] },
\end{equation}
where $g$ is the determinant of the metric, $\kappa=\frac{8\pi G}{c^4}$  where ${ G}$ is the Newtonian gravitational constant and ${ c}$ is the speed of light and ${L}_m$ refers to the Lagrangian of the matter. The variation of the action (\ref{action}) with respect to the metric yields the following field equations:
\begin{equation}\label{FielEq}
 { f_R R_{\mu\nu} - \dfrac{1}{2}g_{\mu\nu}f - \nabla_\mu\nabla_\nu f_R + g_{\mu\nu}\square f_R = \kappa \mathfrak{T}_{\mu\nu}},
\end{equation}
where ${ f_R \equiv df(R)/dR}$, the d'Alembert operator ${\square \equiv \nabla_\mu\nabla^\mu}$ with ${ \nabla_\mu}$ is the covariant derivative associated to Levi-Civita connection and $\mathfrak{T}_{\mu\nu}$ is the matter stress-energy tensor. We note that the equations of motion of nonlinear forms of ${f(R)}$ theories of gravity are of fourth order while the Einstein field equations (second order) are recovered by taking ${f(R)= R}$.

As in the GR theory, the left-hand side of Eq. \eqref{FielEq} is governed by spacetime symmetry. In the present study, we assume the non-rotating spherically symmetric spacetime configuration to govern stellar structure. Thus, we write the line element
\begin{equation}\label{MetricEq}
{    ds^2 = g_{\mu \nu} dx^\mu dx^\nu=-e^{\psi}c^2dt^2 + e^{\lambda}dr^2 + r^2(d\theta^2 + \sin^2\theta d\phi^2)},
\end{equation}
where $x^\mu$ are the four-position vector components: time, $t$, radial distance, $r$, and angular coordinates $\theta$ and $\phi$. The metric functions, denoted as $\psi$ and $\lambda$, depend on the radial coordinate $r$ only. Consequently, we determine $\sqrt{-g}=e^{\psi + \lambda} r^2 \sin \theta$, and the four-velocity vector $v^\mu=(ce^{-\psi}, 0, 0, 0)$.

On the other hand, for the right-hand side of Eq. \eqref{FielEq}, we assume the matter sector to be represented by anisotropic fluid, i.e.
\begin{equation}\label{Tmn-anisotropy}
    { \mathfrak{T}{^\mu}{_\nu}}=  {(p_t+\rho c^2)v{^\mu} v{_\nu}+p_t \delta ^\mu _\nu + (p_{r}-p_t) w{^\mu} w{_\nu}}\,,
\end{equation}
where ${v_\mu}$ and ${w{^\mu}}$ are time-like four-velocity and unit vector in the radial direction, ${\rho=\rho(r)}$ denotes the fluid density, ${p_{r}}={ p_{r}(r)}$ represents radial pressure, i.e. $\parallel { v_\mu}$, and ${ p_t}={ p_t(r)}$ represents tangential pressure, i.e. $\perp w{^\mu}$. Then the stress-energy tensor is given by the diagonal matrix: \[{ \mathfrak{T}{^\mu}{_\nu}}={diag(-\rho c^2\, ,p_{r}\, ,p_t\, ,p_t)}.\]
In $f(R)$ gravity Ricci scalar can be related to the trace of the stress-energy tensor $\mathfrak{T}$ by a second-order differential equation by taking the trace of Eq.~(\ref{FielEq}),
\begin{equation}\label{TraceEq}
  {   3\square f_R(R) + Rf_R(R) - 2f(R) =\kappa \mathfrak{T}}.
\end{equation}
In this sense, the ${f(R)}$ gravity involves $g_{\mu\nu}$ and ${R}$ as dynamical field variables as indicated by the differential equation \eqref{TraceEq}. This is in contrast to the GR scenario which relates Ricci scalar to the matter trace by a simple algebraic equation $R=\kappa \mathfrak{T}$, and subsequently obtains ${R}=0$ whereas no matter present.  In other words, non-linear $f(R)$ could result in a non-zero Ricci scalar even in the outer region of a dense star where is no matter, i.e. $\mathfrak{T}=0$.

As in the GR theory, the matter divergence-free energy-momentum tensor provides the conservation law of energy and momentum. For the spacetime \eqref{MetricEq} and the fluid \eqref{Tmn-anisotropy} we write the continuity equation
\begin{equation}\label{ConservationEq}
 {    \nabla_\mu \mathfrak{T}_r^{\ \mu} = \frac{1}3(p_r+2p_t)' + (\rho + \frac{1}3[p_r+p_t])\psi'= 0 },
\end{equation}
where $'\equiv d/dr$. We further write the field equations \eqref{FielEq} as follows
\begin{widetext}
\begin{eqnarray}
    &&{ -\frac{f_R}{r^2} + \frac{f_R}{r^2}\frac{d}{dr}\left( re^{-2\lambda} \right) + \frac{1}{2}(Rf_R - f) + \frac{1}{e^{2\lambda}}\left[ \left( \frac{2}{r} - \lambda' \right)f_R' + f_R^{''} \right] = -\kappa c^2 \rho,} \label{FielEq1}
\\
&&{ -\frac{f_R}{r^2} + \frac{f_R}{e^{2\lambda}}\left( \frac{2\psi'}{r} + \frac{1}{r^2} \right) + \frac{1}{2}(Rf_R- f) + \frac{1}{e^{2\lambda}}\left( \frac{2}{r} + \psi' \right)f_R' = \kappa  p_r,} \label{FielEq2}
\\
&&{ \frac{f_R}{r^2}\left[ \frac{1}{e^{2\lambda}}(r\lambda' - r\psi' -1) +1 \right] - \frac{1}{2}f + \frac{1}{e^{2\lambda}}\left[ \left( \frac{1}{r} + \psi' - \lambda' \right)f_R' + f_R^{''} \right] = \kappa p_t.} \label{FielEq3}
\end{eqnarray}
\end{widetext}
The above equations reduce to the GR case when $f(R)=R$ \cite{Ray:2003gt}. We additionally note that the equation which governs the dynamics of Ricci scalar, namely (\ref{TraceEq}), takes the following form
\begin{align}\label{RicciScalarEq}
{     \frac{3}{e^{2\lambda}}\left[ \left( \frac{2}{r} + \psi' - \lambda' \right)f_R' + f_R^{''} \right]} &{ + Rf_R - 2f  =\kappa (-\rho c^2+ p_r+2p_t) }.
\end{align}
By now we completed our review of the basic equations of $f(R)$ modified gravity for static spherically symmetric spacetime whereas the matter fluid is anisotropic. In the following we setup a particular stellar model along with a specific $f(R)$ theory.
\section{The Starobinsky $f(R)$ gravity}\label{Sec:Model}

We use a specific $f(R) = R + \epsilon R^2$ theory which includes a quadratic correction of Ricci scalar, namely Starobinsky $f(R)$ gravity, where $\epsilon$ is a dimensionful parameter with [L$^2$]. By substituting into Eqs. \eqref{FielEq1}--\eqref{FielEq3}, we obtain
\begin{align}
    {  \rho} &{ \equiv \dfrac{1}{c^2\kappa}\left\lbrace \dfrac{1+2\epsilon R}{r^2}\dfrac{d}{dr}\bigg[ r\left( 1- e^{-2\lambda} \right)\bigg]-\dfrac{\epsilon}{2}R^2 - \dfrac{2\epsilon}{e^{2\lambda}}\left[ \left( \dfrac{2}{r}- \lambda' \right)R' + R'' \right] \right\rbrace} ,  \label{9}  \\
{ p_r} &{ \equiv \dfrac{1}{\kappa}\left\lbrace -\dfrac{1+2\epsilon R}{r}\left[ \dfrac{2\psi'}{e^{2\lambda}} - \dfrac{1}{r}\left( 1-e^{-2\lambda} \right) \right] + \dfrac{\epsilon}{2}R^2 +\dfrac{2\epsilon}{e^{2\lambda}}\left( \dfrac{2}{r} + \psi' \right)R'  \right\rbrace} ,   \label{10}  \\
{ p_t} & { \equiv \dfrac{1}{\kappa}\left\lbrace\dfrac{1+2\epsilon R}{r^2} \bigg[ 1 + \frac{r\lambda' - r\psi' -1}{e^{2\lambda}} \bigg]- \dfrac{R+\epsilon\,R^2}{2}+ \dfrac{2\epsilon}{e^{2\lambda}}\left[ \left( \dfrac{1}{r}+ \psi' - \lambda' \right)R' + R'' \right] \right\rbrace} .  \label{11}
\end{align}
It is easy to see that the matter density and pressures are modified and reduced to the GR solution \cite{Nashed:2020kjh,Roupas:2020mvs} by accounting for $\epsilon=0$. Clearly, the above system has five unknown functions, and hence two constraints must be assumed to close the system. Those could be EoS to relate the radial and tangential pressures to the density, i.e. $p_r(\rho)$ and $p_t(\rho)$. Another way is to assume reasonable forms of the metric potentials $\lambda(r)$ and $\psi(r)$. We follow the latter approach by assuming Krori-Barua (KB) ansatz to describe the spacetime inside the stellar model.
\subsection{Krori-Barua model}\label{Sec:KB}
We introduce the KB metric potentials \cite{Krori1975ASS}
\begin{equation}\label{eq:KB}
   {  \psi(r)=a_0 (r/R_s)^2+a_1,\,  \qquad \lambda(r)=a_2 (r/R_s)^2}\,.
\end{equation}
where $R_s$ denotes the star radius and the set of the dimensionless parameters $\{a_0, a_1, a_2\}$ can be determined by matching conditions. These forms ensure that the solution is regular everywhere inside the stellar structure. Indeed KB ansatz has been applied in many modified gravity theories and in GR as well. However, in this study, we use the observational constraints on mass and radius of the pulsar J0740+6620 as recently studied by NICER to estimate the model parameter $\epsilon$. It is convenient to introduce the dimensionless parameter $\epsilon_1=\epsilon/\ell^2$ where the length $\ell$ is taken as the radius of a canonical neutron star, i.e. $\ell=10$ km.
Using \eqref{eq:KB} and Eqs \eqref{9}--\eqref{11}, we write
\begin{align}\label{sol}
&  \rho={\frac{1}{{{R_s}}^{8}c^2 \kappa { e^{{\frac {2a_2{r}^{2}}{{{R_s}}^{2}}}}}{r}^{4}}}\bigg\{ {{R_s}}^{8} \left( {r}^{2}+2\epsilon_1\ell^{2} \right) {e^{{\frac {2a_2{r}^{2}}{{{R_s}}^{2}}}}}+ {{R_s}}^{6}\left( 2{r}^{4}a_2-12\epsilon_1{{R_s}}^{2}\ell^2-{{R_s}}^{2}{r }^{2} \right) {e^{{\frac {a_2{r}^{2}}{{{R_s}}^{2}}}}}+2 \bigg[ a_0 \left( a_2-a_0 \right)  \left( 3 a_2+a_0 \right)  \nonumber\\
& \times\left(a_0 -4a_2 \right) {r}^{8 } -2 \left( 12{a_2}^{3}+{a_0}^{3}+13{a_0}^{2}a_2-34a_0\,{a_2}^{2} \right) {{R_s}}^{2}{r}^{ 6}+{{R_s}}^{4} \left(40{a_2}^{2}+3{a_0}^{2}- 48a_0a_2 \right) {r}^{4}+4a_2\,{r}^{2}{{R_s}}^{6}+5{ {R_s}}^{8} \bigg]\ell^2 \epsilon_1 \bigg\} \,, \nonumber\\
& { p_r}=\frac{1}{{{R_s}}^{8}\kappa  {e^{{\frac {2a_2{r}^ {2}}{{{R_s}}^{2}}}}}{r}^{4}}\bigg\{ {{R_s}}^{6}\left(2{r}^{4}a_0 -12\epsilon_1{{R_s}}^{2}\ell^{2}+{{R_s}}^{2}{r }^{2} \right) {e^{{\frac {a_2{r}^{2}}{{{R_s}}^{2}}}}}-{{R_s}}^{8} \left( {r}^{2}+2\epsilon_1 \ell^{2} \right){e^{{\frac {2a_2{r}^{2}}{{{R_s}}^{2}}}}}+2\bigg[7 {{R_s}}^{8}+8a_2{r}^{2}{{R_s}}^{6} \nonumber\\
&-{r}^{4} \bigg(12{a_2}^{2}+11{a_0}^{2}-32a_0a_2 \bigg) {{R_s}}^{4}-2{r}^{6}a_0\left( {a_0}^{2}+6{a_2}^{2}-9a_0a_2 \right) {{R_s}}^{2}+{r}^{8}{a_0}^{2} \left( a_0 -a_2\right)  \left( 3a_2+a_0 \right)  \bigg]\ell^2\epsilon_1 \bigg\}\,,
\nonumber\\
& {  p_t}=\frac{1}{ {{R_s}}^{8}\kappa  {e^{{\frac {2a_2\,{r}^{2}}{{{R_s}}^{2}}}}}{r}^{4}}\bigg\{ 2 {e^{{\frac {2a_2{r}^{2}}{{{R_s}}^{2}}}}}{{R_s}}^{8}\ell^2\epsilon_1+{{R_s}}^{4} \left( a_0 \left(a_0  -a_2\right) {r}^{6}+ \left(2a_0 -a_2 \right) {{R_s}}^{2}{r}^{4}-12{{R_s}}^{2}\ell^2\epsilon_1\left( a_0 -a_2\right) {r}^{2}+12{{R_s}}^{4}\ell^2\epsilon_1 \right) {e^{{ \frac {a_2{r}^{2}}{{{R_s}}^{2}}}}}\nonumber\\
&-2\ell^2\epsilon_1 \bigg[a_0  \left( a_0 -a_2\right) \left( 12a_2^2 -5a_0a_2+a_0^2 \right) {r}^{8}+2 R_s^2 \left( 37a_0 a_2^2-25a_0^2a_2-12a_2^3+4a_0^3 \right) {r}^{6}+ \left(28{a_2}^{2}-38a_0a_2+11{a_0}^{2} \right)R_s^{4}{r}^ {4}\nonumber\\
& -2 \left( 3a_0-7a_2 \right) {{R_s}}^{6}{r}^{2}+7{{R_s}}^{8} \bigg]  \bigg\}.
\end{align}
Additionally, we define the anisotropic force ${ F_a=\frac{2\Delta}{r}}$ which is induced by the pressure difference, equivalently the anisotropy parameter $\Delta= {  p_t}- {  p_{r}}$. Notably the anisotropic force vanishes at the center. For $0<r \leq R_s$, the strong anisotropy case, $\Delta>0$, requires $p_t> p_r$ everywhere inside the star. On the contrary, the mild anisotropy case, $\Delta<0$, requires $p_r> p_t$ everywhere inside the star.

\subsection{Matching conditions}\label{Sec:Match}

Since the vacuum solutions of GR and Starobinsky $f(R)$ gravity are equivalent \citep{Ganguly:2013taa}, the exterior solution is nothing but the Schwarzschild vacuum solution. Therefore, we use the following form of the exterior spacetime
\begin{equation}
 {   ds^2=-\left(1-\frac{2GM}{c^2r}\right) c^2 dt^2+\frac{dr^2}{\left(1-\frac{2GM}{c^2 r}\right)}+r^2 (d\theta^2+\sin^2 \theta d\phi^2)}\,,
\end{equation}
where $M$ is the gravitational mass of the star as measured by an observer at infinite distance. Using the interior spacetime given by Eq. \eqref{MetricEq}, we take the matching conditions at the boundary surface
\begin{equation}\label{eq:bo}
 \psi(r={R_s})=\ln(1-C),  \lambda(r={R_s})=-\ln(1-C)\,  \text{and} \, {{p}_r}(r={R_s})=0,
\end{equation}
where the compactness parameter is defined as
\begin{align}\label{comp11}
 {  C=\frac{2GM}{c^2 {R_s}}}\,.
\end{align}
Using the KB ansatz \eqref{eq:KB} and the radial pressure in Eq. \eqref{sol} along with the above boundary conditions, one can obtain the model parameters $\{a_0, a_1, a_2\}$ in terms of $\epsilon_1$ and the compactness parameters. Therefore, the parameter space for the quadratic $f(R)$ gravity in the present model is in principal \{$\epsilon_1$, $C$\}. Since the mass and radius of pulsars are constrained by astrophysical observations and consequently the compactness, it remains to determine the corresponding constraints on the model parameter $\epsilon_1$.
\section{Astrophysical and stability constraints from the pulsar J0740+6620 observations}\label{Sec:Stability}

In this section, we use observational constraints on the mass and radius of the pulsar J0740+6620 in particular to estimate the value of the quadratic gravity correction parameter $\epsilon_1$. Also, we examine the stability of the obtained solution via several physical constraints. As we have mentioned in the introduction that the accurate observational data plays an extremely important role to constrain the parameter space of possible model. First, we discuss our choice of the PSR J0740+6620 to constrain the quadratic $f(R)$ gravity. The relativistic Shapiro time delay determines the pulsar mass $M= 2.08 \pm 0.07 M_\odot$ \citep{NANOGrav:2019jur,Fonseca:2021wxt}. Interestingly, the pulsar mass approaches the upper limit of an NS at which modifications of gravitational effects are expected to be important. On the other hand, the mass is determined with high precision independent from inclination, since this pulsar is in a binary system. Moreover, by combining the X-ray Multi-Mirror (XMM) Newton dataset, it enhances the low NICER count rate, this determines the pulsar radius $R= 13.7_{-1.5}^{+2.6}$ km \cite{Miller:2021qha}. Another independent analysis obtains $R=12.39_{-0.98}^{+1.30}$ km \cite{Riley:2021pdl}. Remarkably, by applying Gaussian process to a nonparameteric EoS approach using NICER+XMM the mass and radius of PSR J0740+6620 are measured as $M=2.07 \pm 0.11 M_\odot$ and $R=12.34^{+1.89}_{-1.67}$ km \citep{Legred:2021hdx}. This latter measurement is in agreement at 1$\sigma$ level with the obtained results in \cite{Landry:2020vaw}. In this sense, the pulsar PSR J0740+6620 provides a perfect laboratory to constrain the parameter space of the quadratic $f(R)$ gravity of the KB model, i.e. the set of parameters \{$\epsilon$, $a_0$, $a_1$, $a_2$\}.

\subsection{The mass and radius observational limits using pulsar J0740+6620}\label{Sec:obs_const}

In the following, we use the accurate measurements of the mass and radius of the PSR J0740+6620 $M=2.07 \pm 0.11 M_\odot$ and radius $R=12.34^{+1.89}_{-1.67}$ km \citep{Legred:2021hdx} which combines NICER+XMM measurements to constrain the quadratic gravity parameter $\epsilon_1$.
\begin{figure*}
\centering
\subfigure[~The Mass function]{\label{Fig:Mass}\includegraphics[scale=0.45]{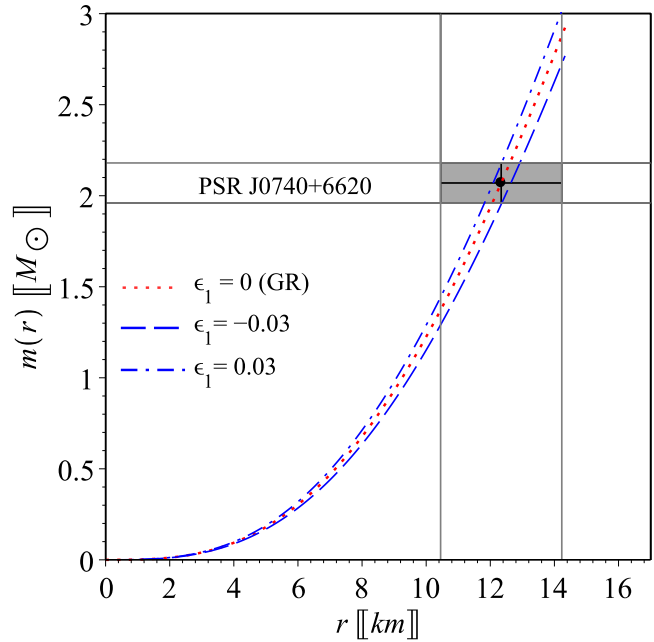}}
\caption{The mass function \eqref{Mf3} for the pulsar J0740+6620: The gray region represents the observational constraints on its mass and radius ($M=2.07\pm 0.11 M_\odot$ and ${R_s}=12.35\pm0.11$ km) \citep{Legred:2021hdx}. For $\epsilon_1=-0.03$, we use model parameters \{$C=0.49557$, $a_0 =0.479$, $a_1 =-1.164$, $a_2 =0.684$\}. For $\epsilon_1=0.03$, we use \{$C=0.56$, $a_0 =0.5$, $a_1 =-1.188$, $a_2 =0.684$\}. For $\epsilon_1=0$ (GR), we use \{$a_0 =0.49122$, $a_1 =-1.1756$, $a_2 =0.684$\}. The plots ensure that the quadratic correction underestimate/overestimate the pulsar mass according to the sign of the model parameter $\epsilon_1$. In all figures in the present study same numerical values of the model parameters are used.}
\label{Fig:Mass1}
\end{figure*}
The matter content within radius, $r$, inside the star is expressed by the mass function
 \begin{align}\label{Mf3}
     {m(r)} =  4\pi\int_{0}^{r} \rho(\zeta) \zeta^2 d\zeta \,.
 \end{align}
Recalling the density profile \eqref{sol}, for Starobinsky gravity $f(R)=R+\epsilon R^2$, we obtain the plots of Fig. \ref{Fig:Mass1} for different values of the parameter $\epsilon_1$.
\begin{itemize}
    \item For $\epsilon_1=0$, the GR case, we fix the numerical values of the set of constants as \{$a_0 =0.49122$, $a_1 =-1.1756$, $a_2 =0.684$\}.
    \item For $\epsilon_1=0.03$, we obtain a gravitational mass $M=2.174 M_\odot$ at a radius of ${R_s} =12.07$ km with a compactness $C=0.56$. This fixes the numerical values of the set of constants as \{$\epsilon_1=0.03$, $C=0.56$, $a_0 =0.500$, $a_1 =-1.184$, $a_2 =0.684$\}.
    \item For $\epsilon_1=-0.03$, we obtain a gravitational mass $M=1.953 M_\odot$ at a radius of ${R_s} =12.88$ km with a compactness $C=0.49557$. This fixes the numerical values of the set of constants as \{$\epsilon_1=-0.03$, $C=0.49557$, $a_0 =0.479$, $a_1 =-1.164$, $a_2 =0.684$\}.
\end{itemize}
This puts constraints on the model parameter $0\leq |\epsilon| \leq 3$ km$^2$. In general, as shown in Fig. \ref{Fig:Mass1}, the quadratic gravity term $R^2$ changes the mass of the pulsar. For $\epsilon>0$, the model predicts mass exceeding the GR, $\epsilon=0$, value at the same radius (smaller size for the same mass). On the contrary, for $\epsilon<0$, the model predicts mass relatively lesser than the GR one (larger size for the same mass). Accordingly, we obtained the changes in the compactness parameter as mentioned above. In the following, we are going to use the above-mentioned numerical values to examine the robustness of the present model against various stability conditions.
\subsection{Geometric sector}\label{Sec:geom}
\begin{figure}
\centering
\subfigure[~Matching solutions]{\label{fig:Junction}\includegraphics[scale=0.38]{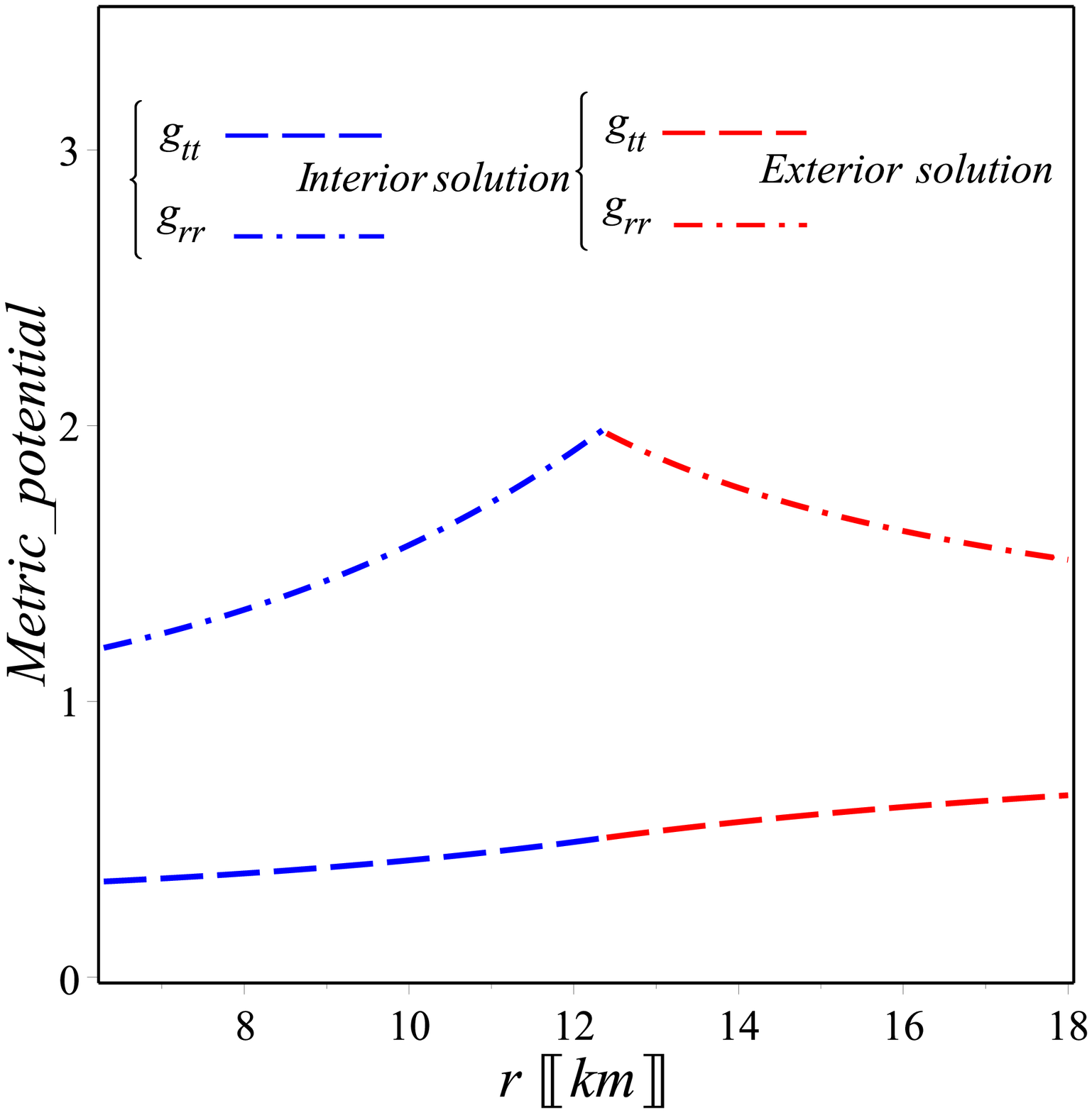}}\hspace{1cm}
\subfigure[~Gravitational redshift]{\label{fig:redshift}\includegraphics[scale=0.38]{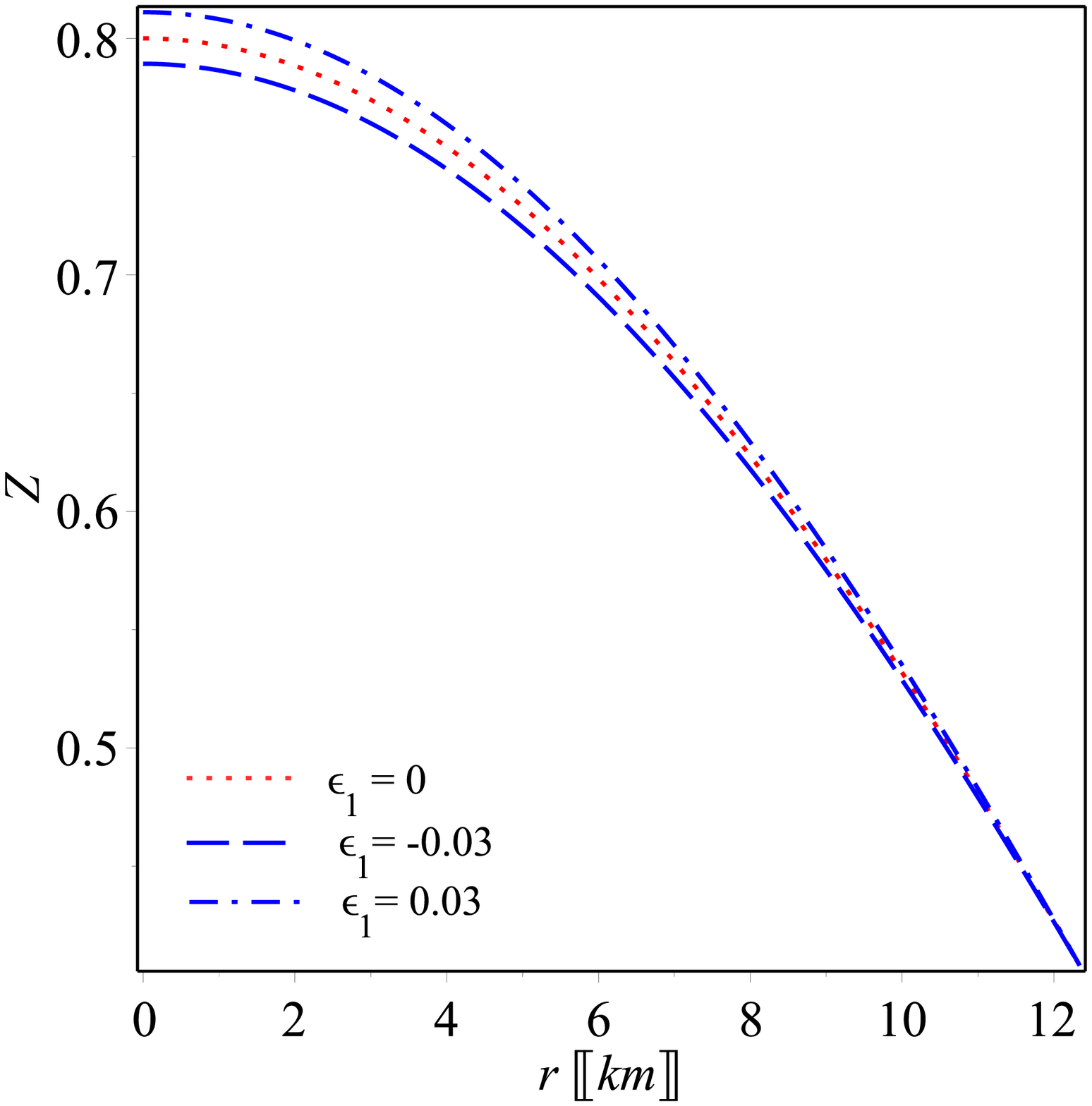}}
\caption{The geometric sector of the pulsar J0740+6620: \subref{fig:Junction} the metric potentials $g_{tt}$ and $g_{rr}$ inside the pulsar J0740+6620 as obtained for KB ansatz and outside the pulsar as obtained by Schwarzschild exterior vacuum solution. The plots ensure that the metric potentials are finit everywhere inside the pulsar and match smoothly with the exterior region. \subref{fig:redshift} the red shift function \eqref{eq:redshift} for $\epsilon_1=0,~ \pm 0.03$ the maximum redshift at the center $Z_s\approx 0.8$ and monotonically decreases to $\approx 0.4$ at the surface of the pulsar for all cases.}
\label{Fig:Matching}
\end{figure}

We note that the metric potentials ${\textit g_{tt}}$ and ${\textit g_{rr}}$ must be nonsingular everywhere inside the star. The KB ansatz \eqref{eq:KB} ensures that the potentials are regular at the center since ${ g_{tt}(r=0)=e^{a_1}}\neq 0$ and ${ g_{rr}(r=0)=1}$, whilst the behavior of the potential at an arbitrary radial distance inside the star is as indicated in Fig. \ref{Fig:Matching} \subref{fig:Junction}. Additionally, we show the matching between the interior KB solution and the exterior (Schwarzschild) one at the boundary of the pulsar J0740+6620, i.e. $R_s=12.34$ km, using our estimated numerical values of the model parameters as mentioned in the figure caption.

Also, we write the gravitational redshift function corresponds to the KB potentials within Starobinsky quadratic $f(R)$ gravity
\begin{equation}\label{eq:redshift}
    Z(r)=\frac{1}{\sqrt{-g_{tt}}}-1=\frac{1}{\sqrt{e^{a_0 (r/R_s)^2+a_1}}}-1.
\end{equation}
We plot the redshift function of the pulsar J0740+6620 for different values of the model parameter $\epsilon_1$ as shown in Fig. \ref{Fig:Matching}\subref{fig:redshift}. For $\epsilon_1=0$ (the GR case), the redshift at the center is $Z(0)\approx 0.8063$ and its value at the surface $Z_{R_s}\approx 0.4$. For $\epsilon_1=- 0.03$, we obtain the maximum redshift at the center $Z(0)\approx 0.794$ (less than the GR value) which monotonically decreases toward the surface where the redshift is $Z_{R_s}\approx 0.4042$ (almost the same as the GR result) less than the upper bound constraint $Z_{R_s}=2$, see \citep{Buchdahl:1959zz,Ivanov:2002xf,Barraco:2003jq,Boehmer:2006ye}. Similarly, for $\epsilon_1= 0.03$, we obtain the maximum redshift at the center $Z(0)\approx 0.811$ (greater than the GR value) and monotonically decreases toward the surface where the redshift is $Z_{R_s}\approx 0.4052$ (almost the same as the GR result) less than the upper bound constraint $Z_{R_s}=2$. In conclusion, in both cases the redshift patterns of the quadratic $f(R)$ gravity satisfy the stability conditions, that is the gravitational redshift is a positive finite value everywhere inside the star and decreases monotonically toward the boundary, i.e. $Z>0$ and $Z'<0$, as shown in Fig. \ref{Fig:Matching}\subref{fig:redshift}.

\subsection{Matter sector}\label{Sec:matt}

\begin{figure*}
\centering
\subfigure[~The energy-density]{\label{fig:density}\includegraphics[scale=0.3]{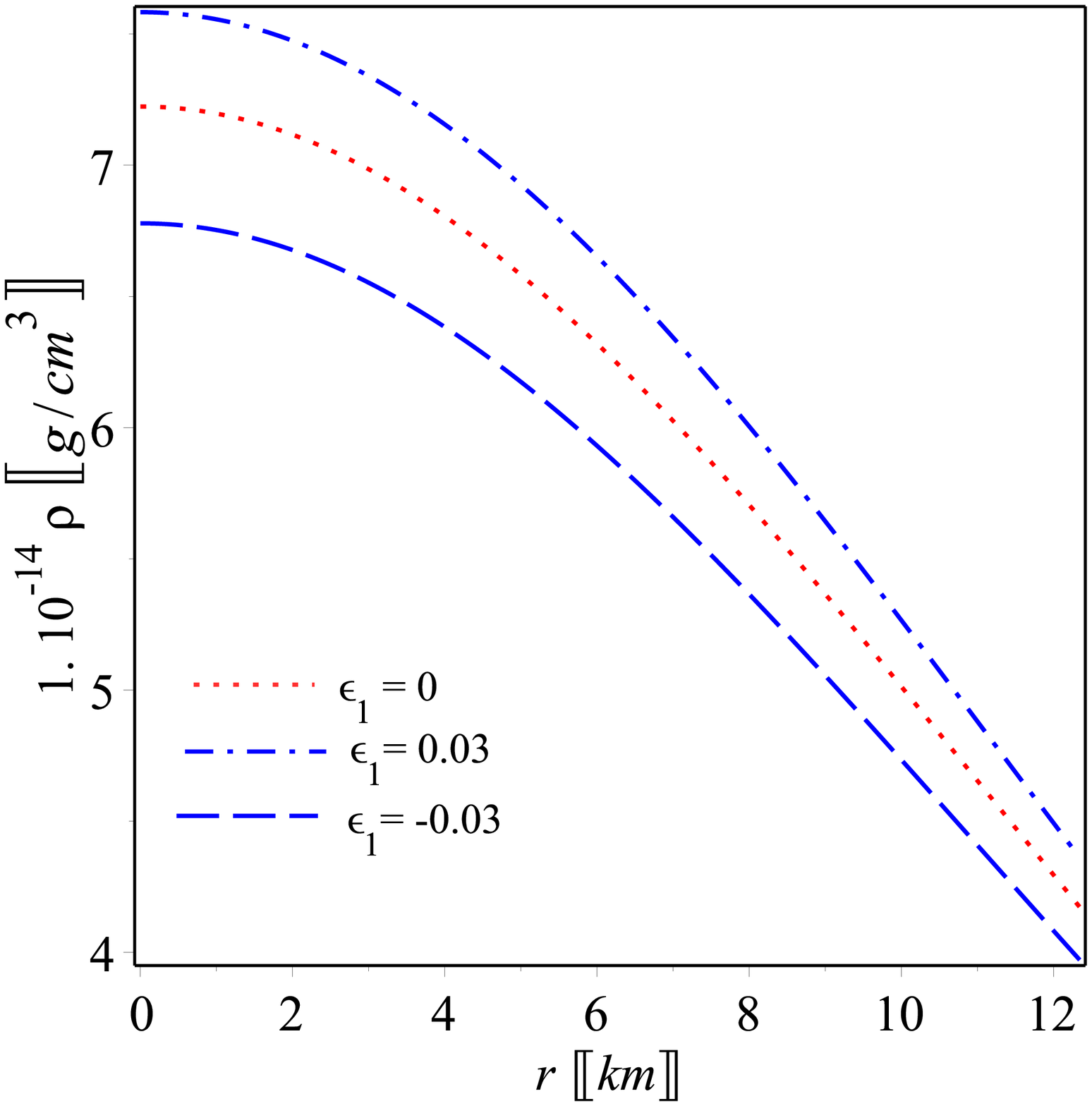}}\hspace{0.5cm}
\subfigure[~The radial pressure]{\label{fig:radpressure}\includegraphics[scale=0.3]{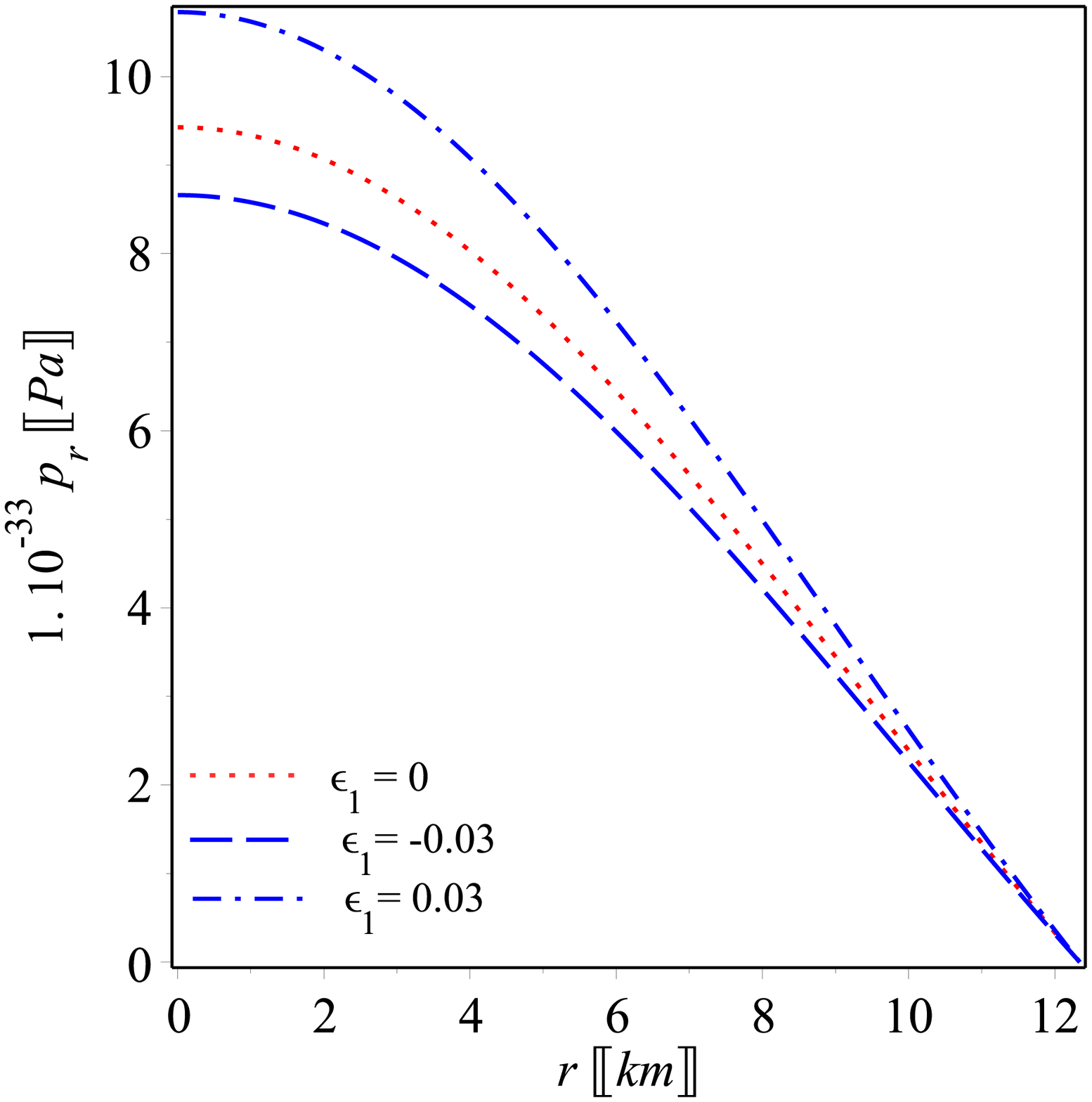}}\\
\subfigure[~The tangential pressure]{\label{fig:tangpressure}\includegraphics[scale=0.3]{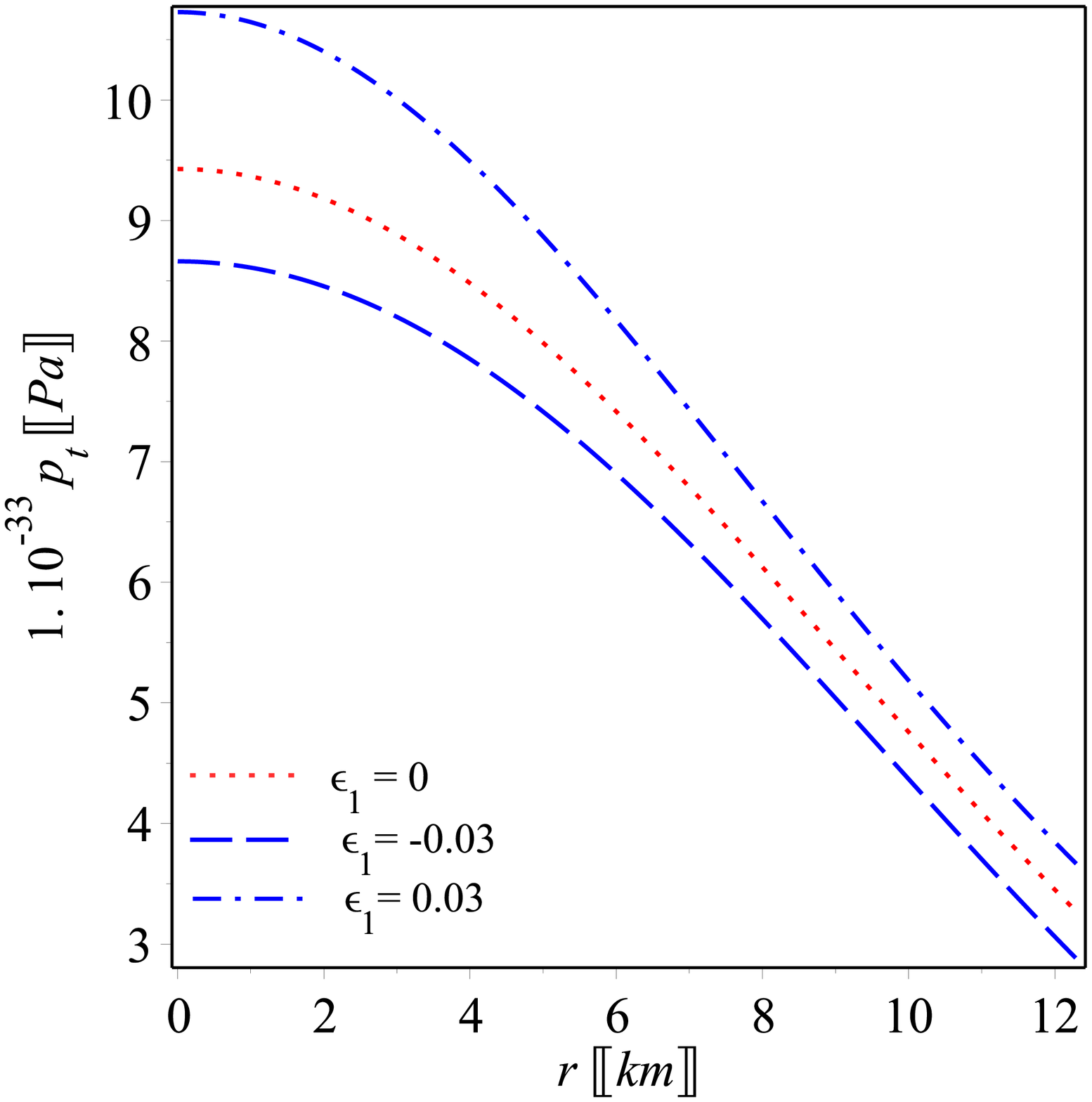}}\hspace{0.5cm}
\subfigure[~The anisotropy function]{\label{fig:anisotf}\includegraphics[scale=0.3]{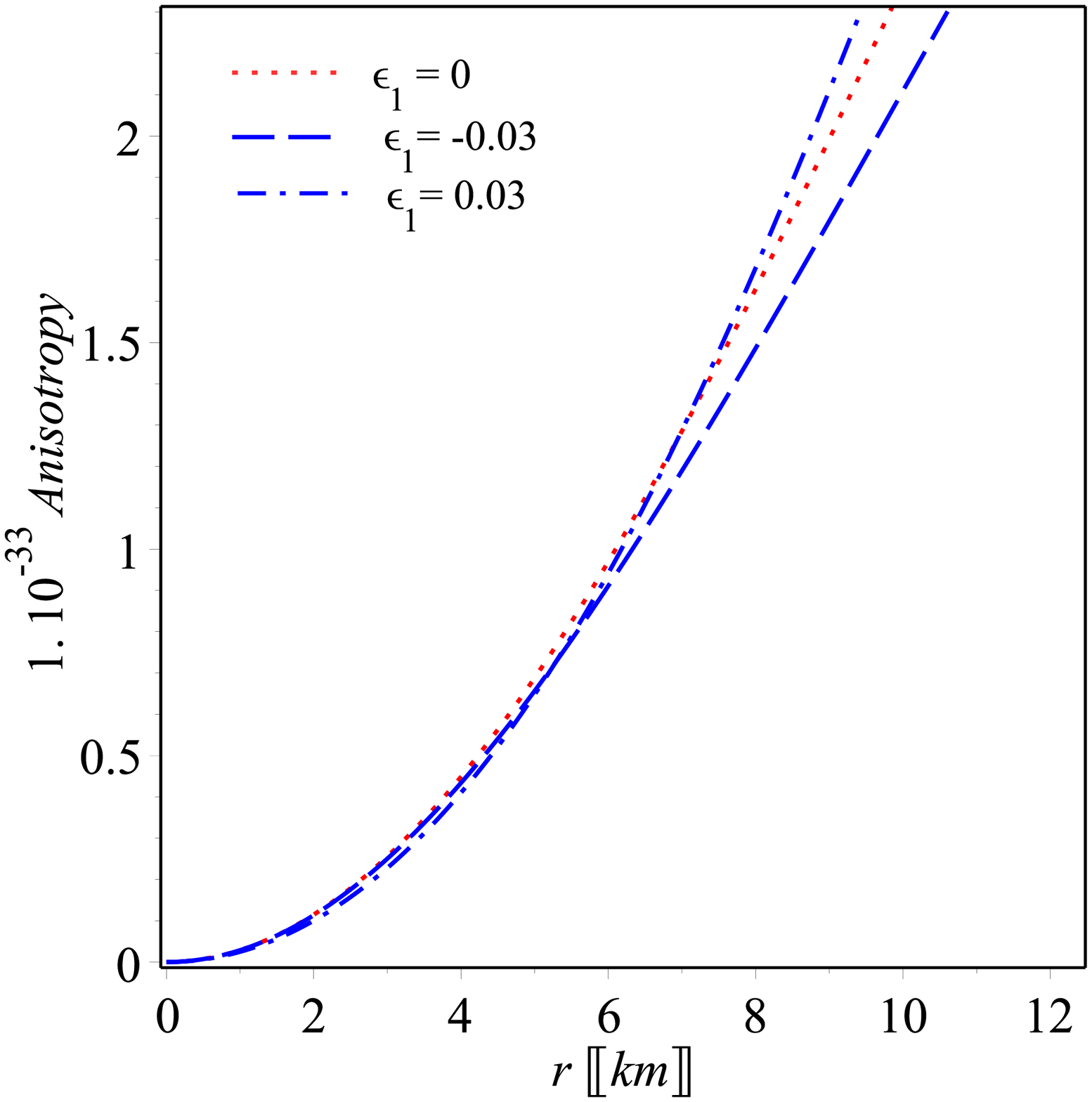}}
\caption{The matter sector of the pulsar J0740+6620: \subref{fig:density}--\subref{fig:tangpressure} represent the density profile and the radial and tangential pressures \eqref{sol} of the pulsar J0740+6620 for $\epsilon_1=0,~\pm 0.03$. The plots ensure that the density and pressures are finite everywhere inside the pulsar and monotonically decrease toward the surface. \subref{fig:anisotf} represents the anisotropy distribution $\Delta(r)=p_t-p_r$ inside the pulsar for $\epsilon_1=0,~\pm 0.03$. Clearly it vanishes at the center as $p_t=p_r$, whereas $\Delta$ is positive (strong anisotropy $p_t>p_r$) elsewhere which induces repulsive force, $F_a=2\Delta/r$, responsible for resizing the pulsar radius.}
\label{Fig:dens_press}
\end{figure*}

Recalling Eqs. \eqref{sol} along with our numerical estimation of the model parameters as given in Subsection \ref{Sec:obs_const}, we obtain the plots of the density, radial and tangential pressures in the radial distance as indicated in Figs. \ref{Fig:dens_press}\subref{fig:density}--\subref{fig:tangpressure}. Obviously, the density and the pressures profiles satisfy the stability conditions related to the matter sector since they are maximum at the center and positive nonsingular everywhere inside the star and monotonically decrease toward the star surface, i.e. $\rho(0)>0$, $\rho'(0)=0$, $\rho''(0)<0$, $\rho(r)>0$, $\rho'(r)>0$ and similarly for the radial and tangential pressures. In addition, we plot the anisotropy $\Delta(r):=p_t-p_r$ as shown in Fig. \ref{Fig:dens_press}\subref{fig:anisotf}. This shows that the anisotropy satisfies the stability condition since it vanishes at the center and increases monotonically toward the surface. We note that the strong anisotropy case, as in the present study, contributes in the hydrodynamic equilibrium with an extra positive force (against gravitational force) proportional to $\Delta/r$ which plays an essential role in resizing the star allowing the star to gain more mass in comparison to the isotropic and mild anisotropic cases. This will be discussed in some detail in Subsection \ref{Sec:TOV}.

It is worth to give some numerical values of some physical quantities of the pulsar J0740+6620 as predicted by the present model as follows: For $\epsilon=0.03$, the core density ${\rho_\text{core}\approx 7.58\times 10^{14}}$ g/cm$^{3} \approx 2.8\rho_\text{nuc}$ and $p_{r\text(core)}\approx 1.03\times 10^{34}~\text{dyn/cm}^2 \approx p_{t\text(core)}$. At the surface we find  ${\rho_s\approx 4.36\times 10^{14}}$ g/cm$^{3} \approx 1.6 \rho_\text{nuc}$, $p_{r(R_s)}\approx 0$ dyn/cm$^2$ and $p_{t(R_s)}=3.52\times 10^{34}$ dyn/cm$^2$. For $\epsilon=-0.03$, the core density ${\rho_\text{core}\approx 6.78\times 10^{14}}$ g/cm$^{3} \approx 2.5\rho_\text{nuc}$ and $p_{r\text(core)}\approx 8.66\times 10^{33}~\text{dyn/cm}^2 \approx p_{t\text(core)}$. At the surface we find  ${\rho_s\approx 3.97\times 10^{14}}$ g/cm$^{3} \approx 1.4\rho_\text{nuc}$, $p_{r(R_s)}\approx 0$ dyn/cm$^2$ and $p_{t(R_s)}\approx 2.85\times 10^{33}$ dyn/cm$^2$. In this sense, the model does not exclude the possibility of the core of the pulsar J0740+6620 being neutrons. Also, the density/pressures values well justify the anisotropy assumption.

As mentioned in Section \ref{Sec:Model} that KB ansatz \eqref{eq:KB} has been used, instead of using EoSs, to close the system \eqref{9}--\eqref{11}. However, we show that the KB ansatz effectively relates the pressures and density. Therefore, we introduce the parameter $\eta:=r/R_s$, then we write the power series of Eqs. \eqref{sol} up to $O(\eta^4)$. It can be shown that those equations induce the following relations
\begin{equation}\label{eq:KB_EoS}
    p_r(\rho)\approx c_1 \rho+c_2\,, \qquad  p_t(\rho) \approx c_3 \rho+c_4\,,
\end{equation}
where the constants, $c_1, ..., c_4$, are completely determined by the model parameter space $\{\epsilon_1,a_0,a_1,a_2\}$ as given in appendix \ref{Sec:App_1}. Interestingly, we can rewrite the above equations in a more physical form
\begin{equation}\label{eq:KB_EoS2}
    p_r(\rho)\approx v_r^2(\rho-\rho_{1})\,, \qquad  p_t(\rho) \approx v_t^2 (\rho-\rho_{2})\,,
\end{equation}
where the sound speed in radial direction $v_r^2=c_1$, the density $\rho_1=-c_2/c_1$, the sound speed in tangential direction $v_t^2=c_3$ and the density $\rho_2=-c_4/c_3$. Notably the density $\rho_1$ is the surface density $\rho_s$ which satisfies the boundary condition $p_r(\rho_s)=0$. This is not applicable for $\rho_2$ since $p_t$ on the surface does not necessarily vanish. Those encompass the maximally compact EoS (for hadron matter) where $v_r^2=c^2$ and MIT bag model EoS (for quark matter) where $v_r^2=c^2/3$ as special cases. The sound speed and the surface density are not free parameters, they however are completely determined by the present model as shown in appindex \ref{Sec:App_1}. For $\epsilon_1=0.03$, by virtue of Eqs. \eqref{eq:A1}--\eqref{eq:A4}, we calculate $v_r^2=c_1\approx 0.41c^2$, $v_t^2=c_3\approx 0.31 c^2$, $\rho_1=\rho_s=-c_2/c_1\approx 4.82\times 10^{14}$ g/cm$^3$ and $\rho_2=-c_4/c_3\approx 3.85\times 10^{14}$ g/cm$^3$. Similarly, for $\epsilon_1=-0.03$, we calculate $v_r^2\approx 0.35c^2$, $v_t^2\approx 0.22c^2$, $\rho_1=\rho_s\approx 4.0\times 10^{14}$ g/cm$^3$ and $\rho_2\approx 2.49 \times 10^{14}$ g/cm$^3$. Although the induced EoSs \eqref{eq:KB_EoS2} are mostly valid at the center of the star as assumed by the power series approximation, the surface density values, for both $\epsilon_1$ cases, are consistent with the obtained exact values. This justifies the validity of those EoSs everywhere inside the star $0\leq \eta<1$. The validity of the EoSs will be examined with the sound speed later on in Subsection \ref{Sec:causality}.
\subsection{Zeldovich condition}
One of the essential conditions to grantee the stability of the star has been given by Zeldovich \citep{1971reas.book.....Z}, that is the radial pressure at the center of the star at most equals to the central energy density, i.e.
\begin{equation}\label{eq:Zel}
    {\frac{{p}_r(0)}{c^2{\rho}(0)}\leq 1.}
\end{equation}
Recalling Eqs. \eqref{sol}, we obtain the central density and radial pressure
\begin{align}
 & {c^2 {\rho}(r\to0) ={\frac {6 \left( 13\,{a_2}^{2}+{a_0}^{2}-16\,a_0\,a_2 \right) \epsilon_1+3a_2}{{R_s}^{2}{ \kappa}^{2}}}}\,, \nonumber \\
 &  {p_r(r\to0) ={\frac { \left( -22{a_0}^{2}-34{a_2}^{2}+ 64\,a_0\,a_2 \right) \epsilon_1+2\, \left( a_0-1/2\,a_2 \right) }{{R_s}^{2}{\kappa}^{2}}}= {p}_{t}(r\to 0)}.\quad
&\end{align}
Using the numerical values as obtained for the pulsar J0740+6620 earlier in Subsection \ref{Sec:obs_const}, for $\epsilon_1=0.03$ the Zeldovich inequality \eqref{eq:Zel} reads $\frac{{p}_r(0)}{c^2{\rho}(0)}=0.15 < 1$, and for $\epsilon_1=-0.03$ the inequality reads $\frac{{p}_r(0)}{c^2{\rho}(0)}=0.14 < 1$. This confirms that the Zeldovich constraint is fulfilled for both cases.

\subsection{Energy conditions}\label{Sec:Energy-conditions}

It proves convenient to write the field equations \eqref{FielEq} in the following form
\begin{equation}\label{eq:fR_MG}
    G_{\mu\nu}=\kappa\left(\mathfrak{T}_{\mu\nu}+\mathfrak{T}_{\mu\nu}^{geom}\right)=\kappa \bar{\mathfrak{T}}_{\mu\nu},
\end{equation}
where $G_{\mu\nu}:=R_{\mu\nu}-g_{\mu\nu}R/2$ is Einstein tensor, the correction due to the counterpart of the $f(R)$ theory \cite{DeFelice:2010aj}
\begin{equation}
    \mathfrak{T}_{\mu\nu}^{geom}=\frac{1}{\kappa}\left(g_{\mu\nu}(f-R)/2+\nabla_\mu \nabla_\nu f_R- g_{\mu\nu} \square f_R+(1-f_R)R_{\mu\nu}\right),
\end{equation}
and the total stress-energy tensor in the mixed form is given by $\bar{\mathfrak{T}}_{\mu}{^\nu}=diag(-\bar{\rho} c^2, \bar{p}_r, \bar{p}_t, \bar{p}_t)$.
\begin{figure*}
\centering
\subfigure[~WEC \& NEC (radial)]{\label{fig:Cond1}\includegraphics[scale=0.27]{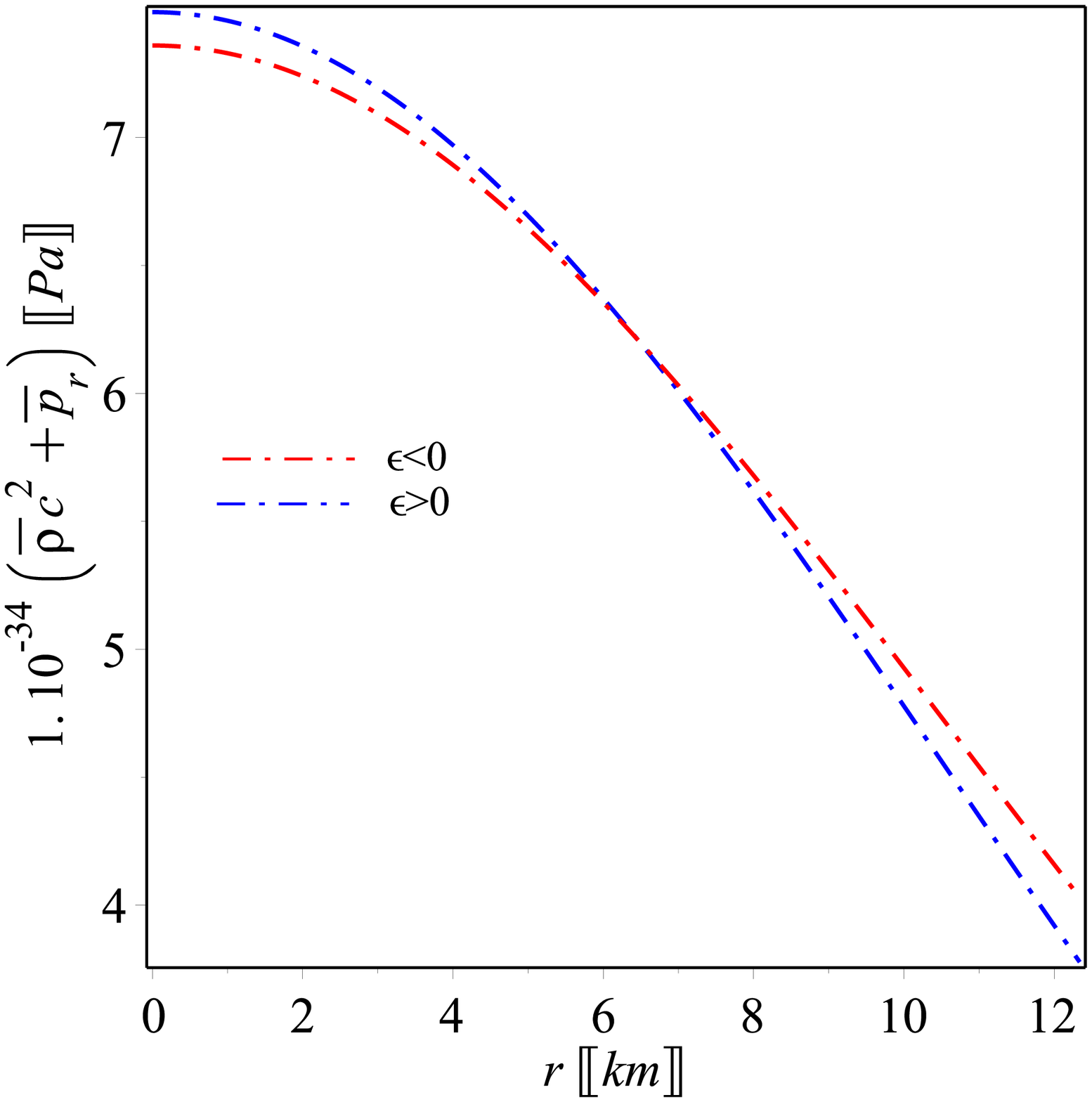}}\hspace{0.2cm}
\subfigure[~WEC \&NEC (tangential)]{\label{fig:Cond2}\includegraphics[scale=0.27]{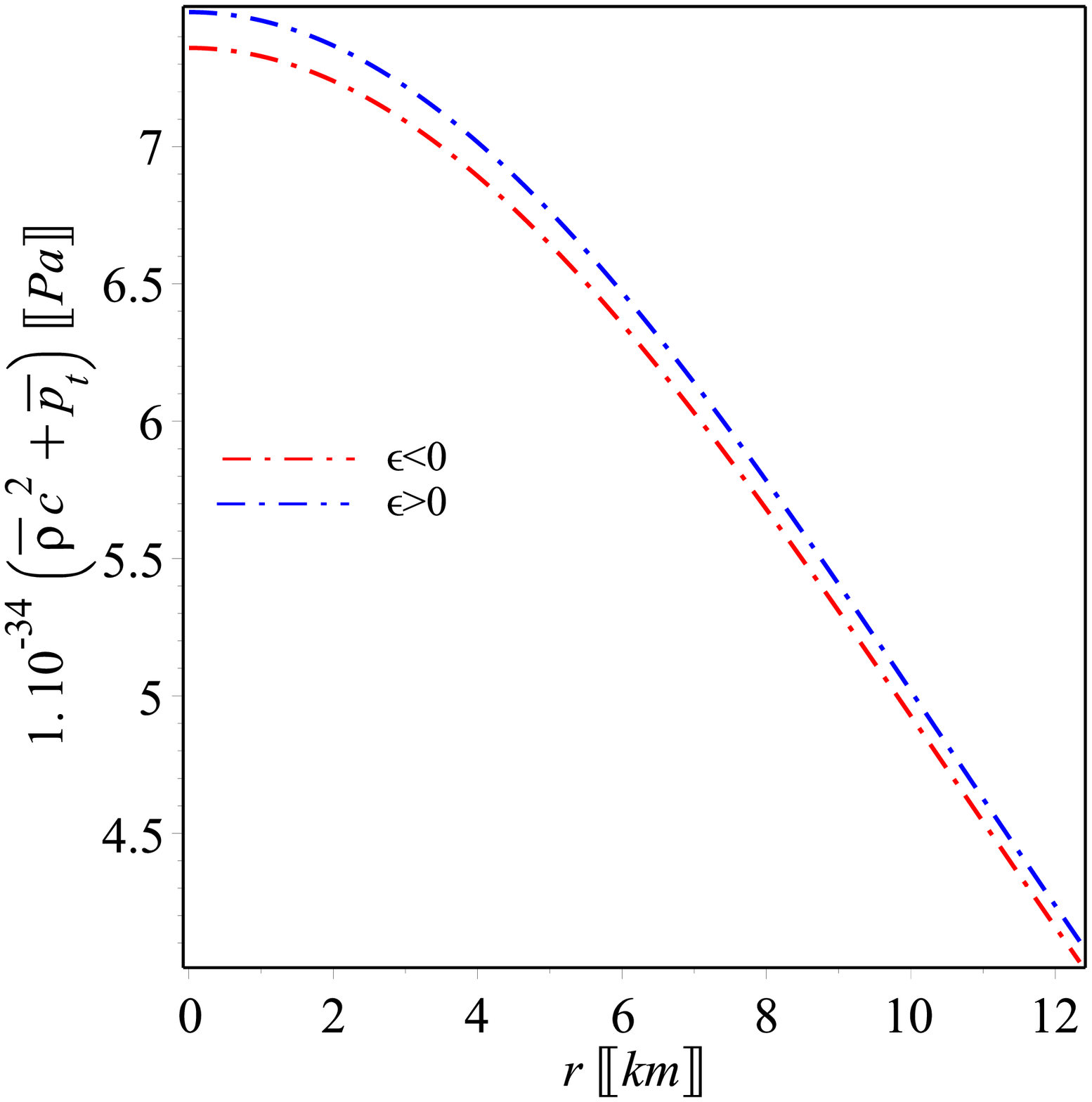}}\hspace{0.2cm}
\subfigure[~SEC]{\label{fig:Cond3}\includegraphics[scale=.27]{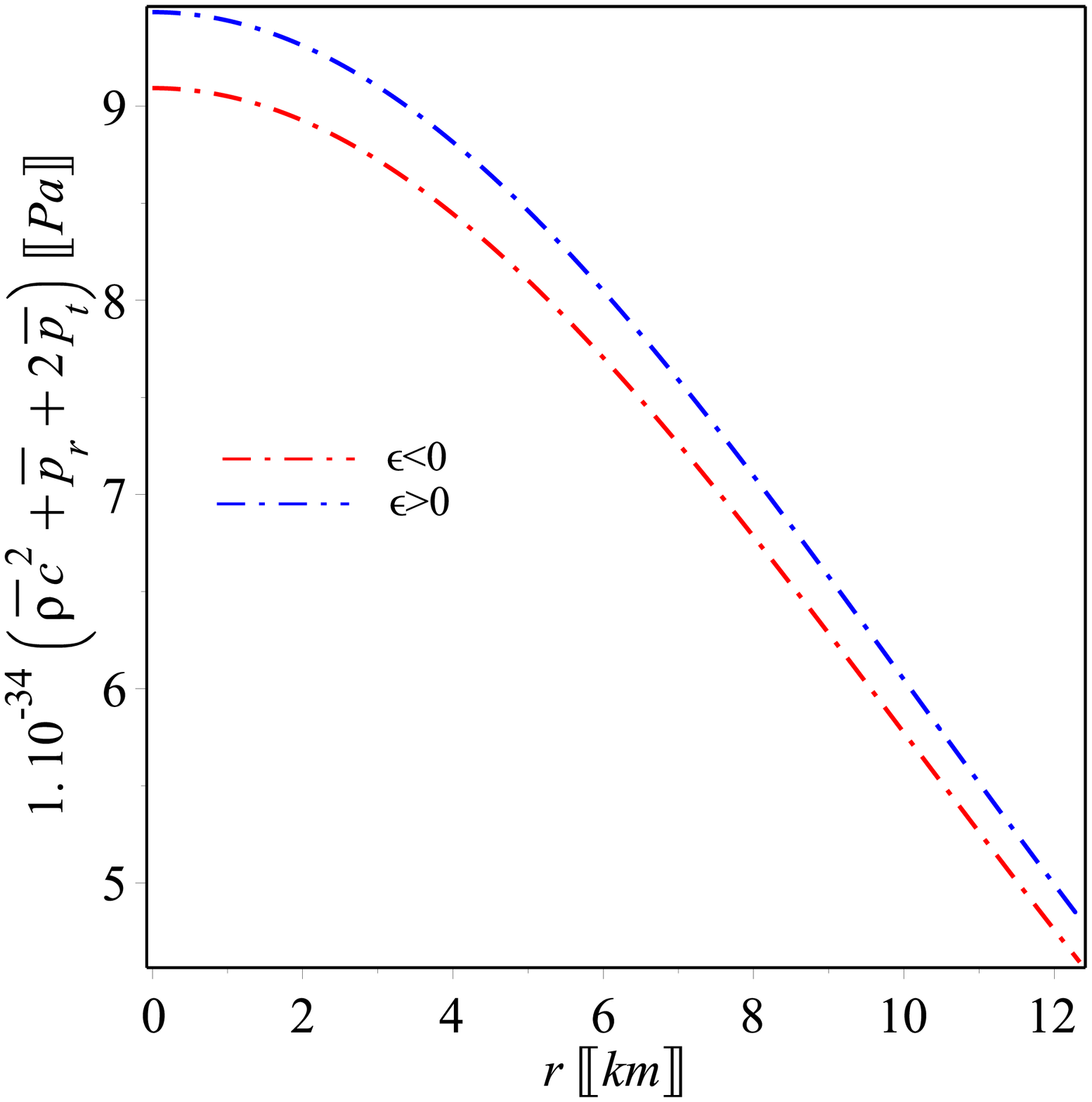}}\\
\subfigure[~DEC (radial)]{\label{fig:DEC}\includegraphics[scale=.27]{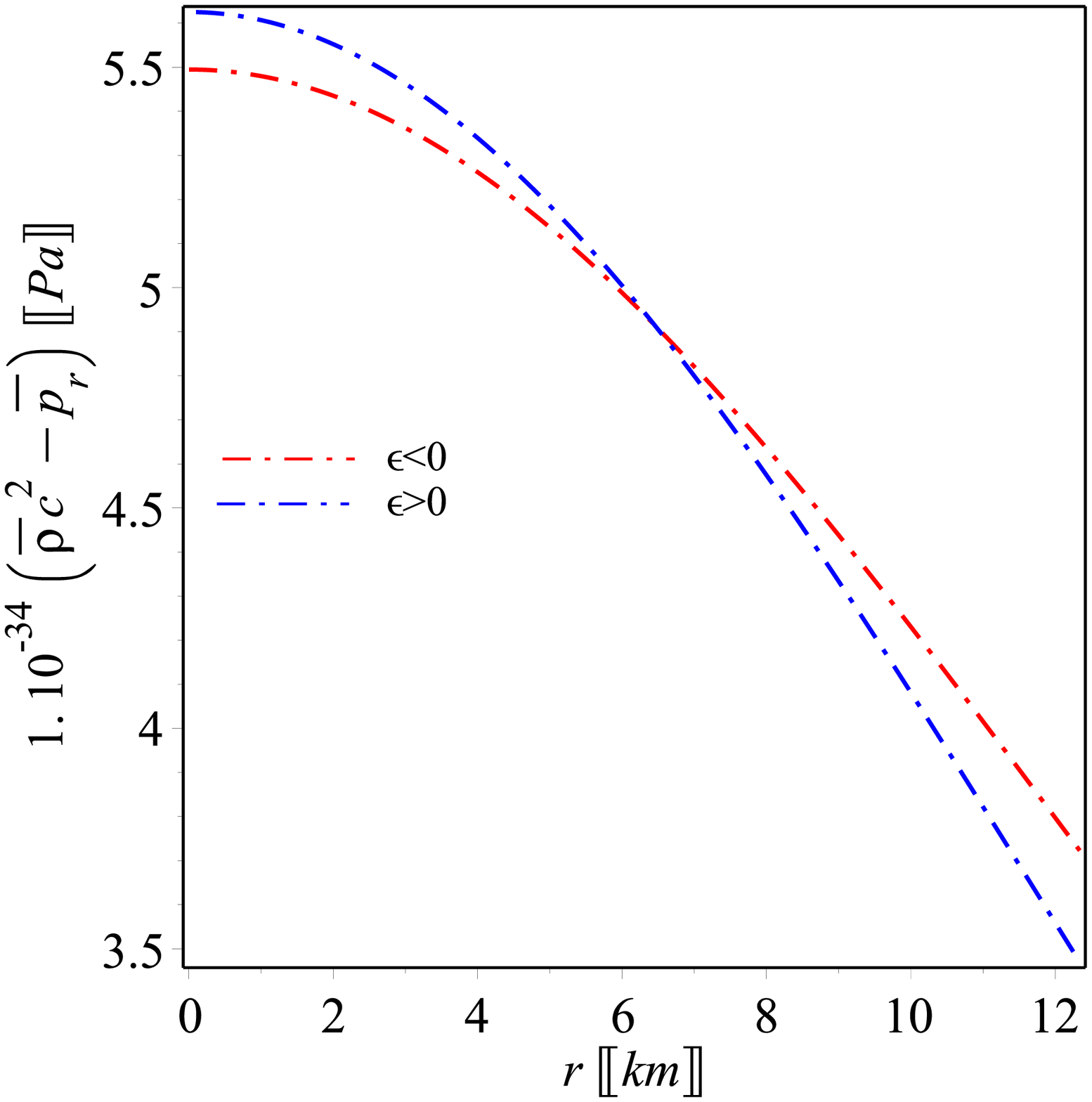}}\hspace{0.2cm}
\subfigure[~DEC (tangential)]{\label{fig:DEC}\includegraphics[scale=.27]{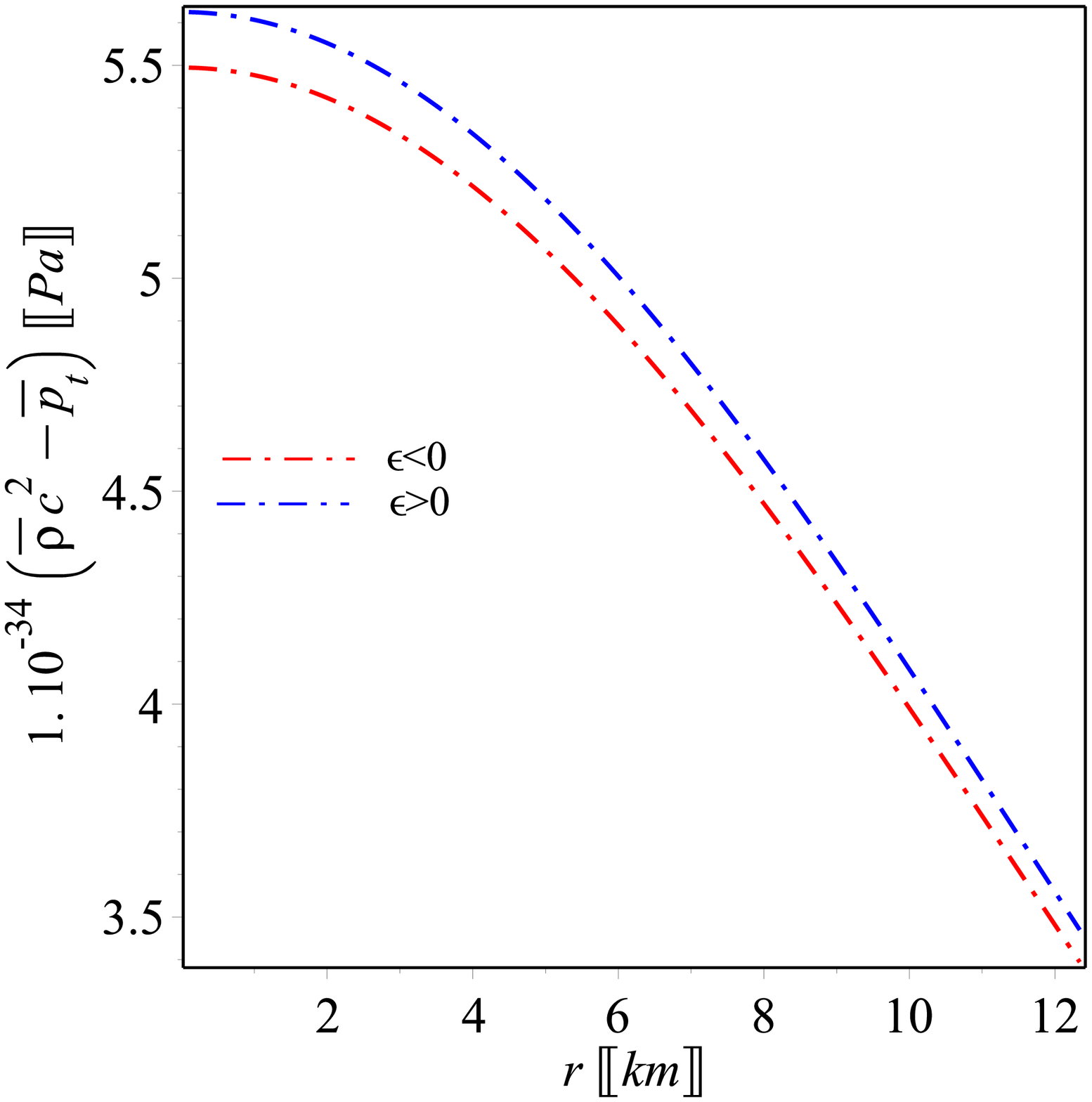}}
\caption{The plots ensure that all the energy conditions for the effective matter $\bar{\mathfrak{T}}_{\mu\nu}$, as discussed in Subsection \ref{Sec:Energy-conditions}, are fulfilled for the obtained model of the pulsar J0740+6620. The cases $\epsilon>0$ and $\epsilon<0$ match $\epsilon_1=0.03$ and $\epsilon_1=-0.03$ respectively.}
\label{Fig:EC}
\end{figure*}

Using Raychaudhuri equation and focusing theorem, it implies that the trace of the tidal tensor to satisfy the following inequalities $R_{\mu\nu} u^{\mu} u^{\nu} \geq 0$ and $R_{\mu\nu} n^{\mu} n^{\nu} \geq 0$, where $u^{\mu}$ is an arbitrary timelike vector and $n^{\mu}$ is an arbitrary future-directed null vector. Keep in mind that Ricci tensor in $f(R)$ modified gravity can be written as $R_{\mu\nu}=\kappa\left(\bar{\mathfrak{T}}_{\mu\nu}-\frac{1}{2} g_{\mu\nu} \bar{\mathfrak{T}}\right)$, as it can be realized by virtue of Eq. \eqref{eq:fR_MG}. In this sense, one could extend the energy conditions to $f(R)$ gravity as follows:
\begin{itemize}
  \item[a.] $\bar{\rho}\geq 0$, $ \bar{\rho} c^2+ \bar{p}_r > 0$ and $\bar{\rho} c^2+\bar{p}_t > 0$, that is the weak energy condition (WEC).
  \item[b.] $\bar{\rho} c^2+ \bar{p}_r \geq 0$ and $\bar{\rho} c^2+  \bar{p}_t \geq 0$, that is the null energy condition (NEC).
  \item[c.] $\bar{\rho} c^2+\bar{p}_r+2\bar{p}_t \geq 0$, $\bar{\rho} c^2+\bar{p}_r \geq 0$ and $\bar{\rho} c^2+\bar{p}_t \geq 0$, that is the strong energy condition (SEC).
  \item[d.] $\bar{\rho}\geq 0$, $\bar{\rho} c^2-\bar{p}_r \geq 0$ and $\bar{\rho} c^2-\bar{p}_t \geq 0$, that is the dominant energy conditions (DEC).\\
\end{itemize}
In Fig. \ref{Fig:EC}, we visualize the energy conditions, for positive and negative values of the model parameter $\epsilon_1$, in terms of the total stress-energy tensor. The plots verify that all energy conditions are fulfilled by the present model of the pulsar J0740+6620.

\subsection{Causality of the model}\label{Sec:causality}

\begin{figure*}
\centering
\subfigure[~Radial speed of sound]{\label{fig:vr}\includegraphics[scale=0.28]{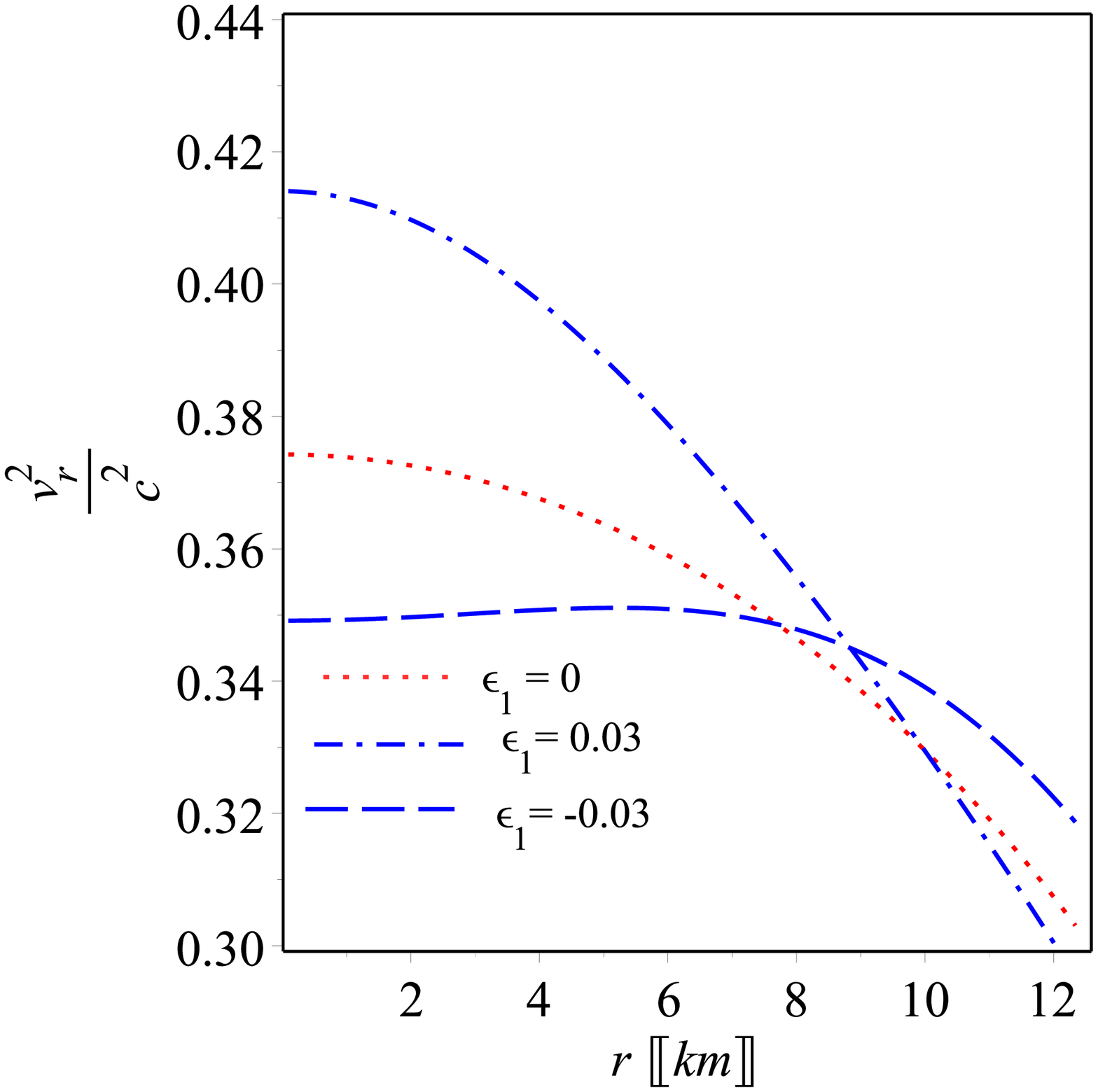}}\hspace{0.2cm}
\subfigure[~Tangential speed of sound]{\label{fig:vt}\includegraphics[scale=.28]{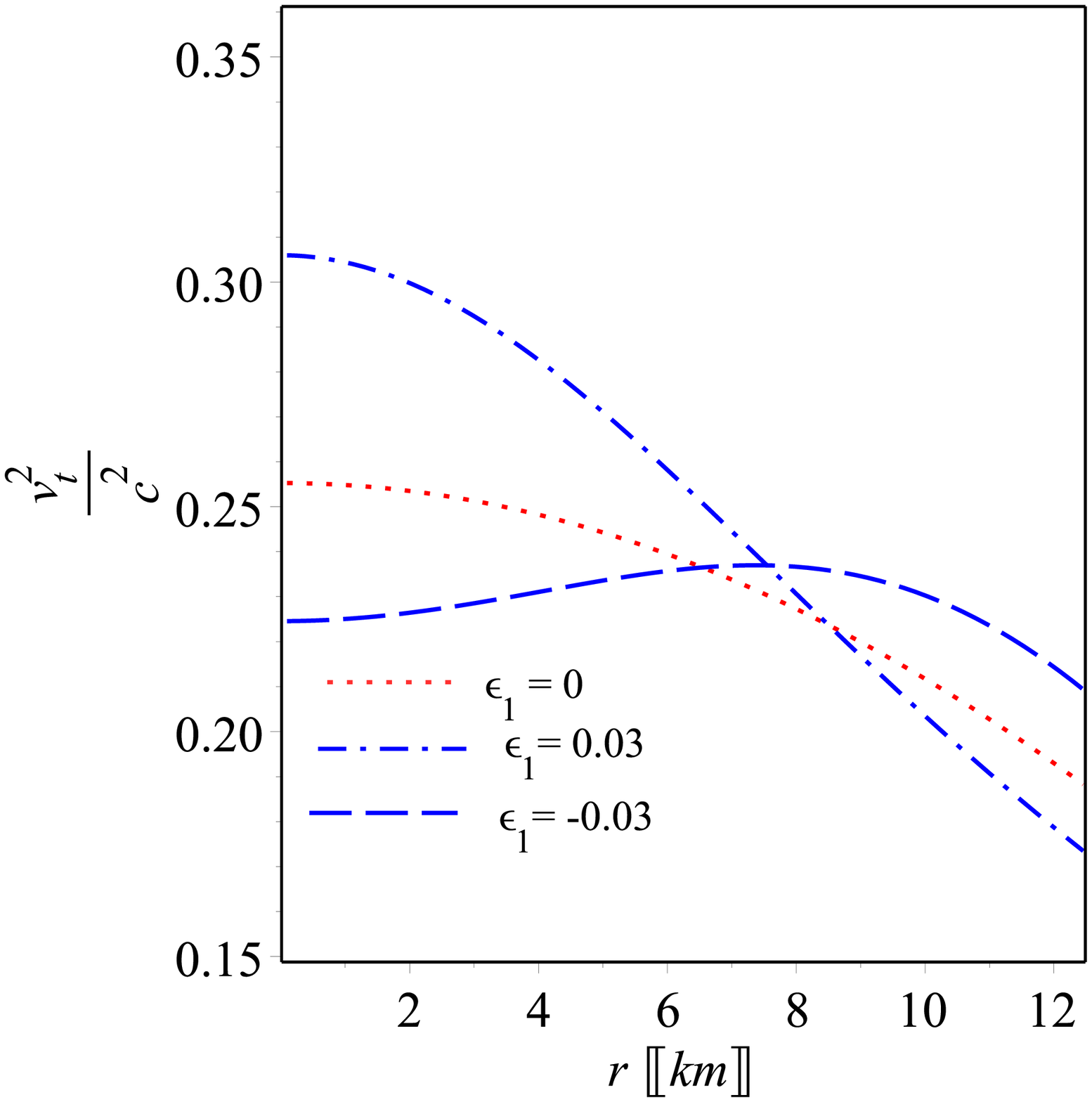}}\hspace{0.2cm}
\subfigure[~Strong anisotropy stability]{\label{fig:vt-vr}\includegraphics[scale=.28]{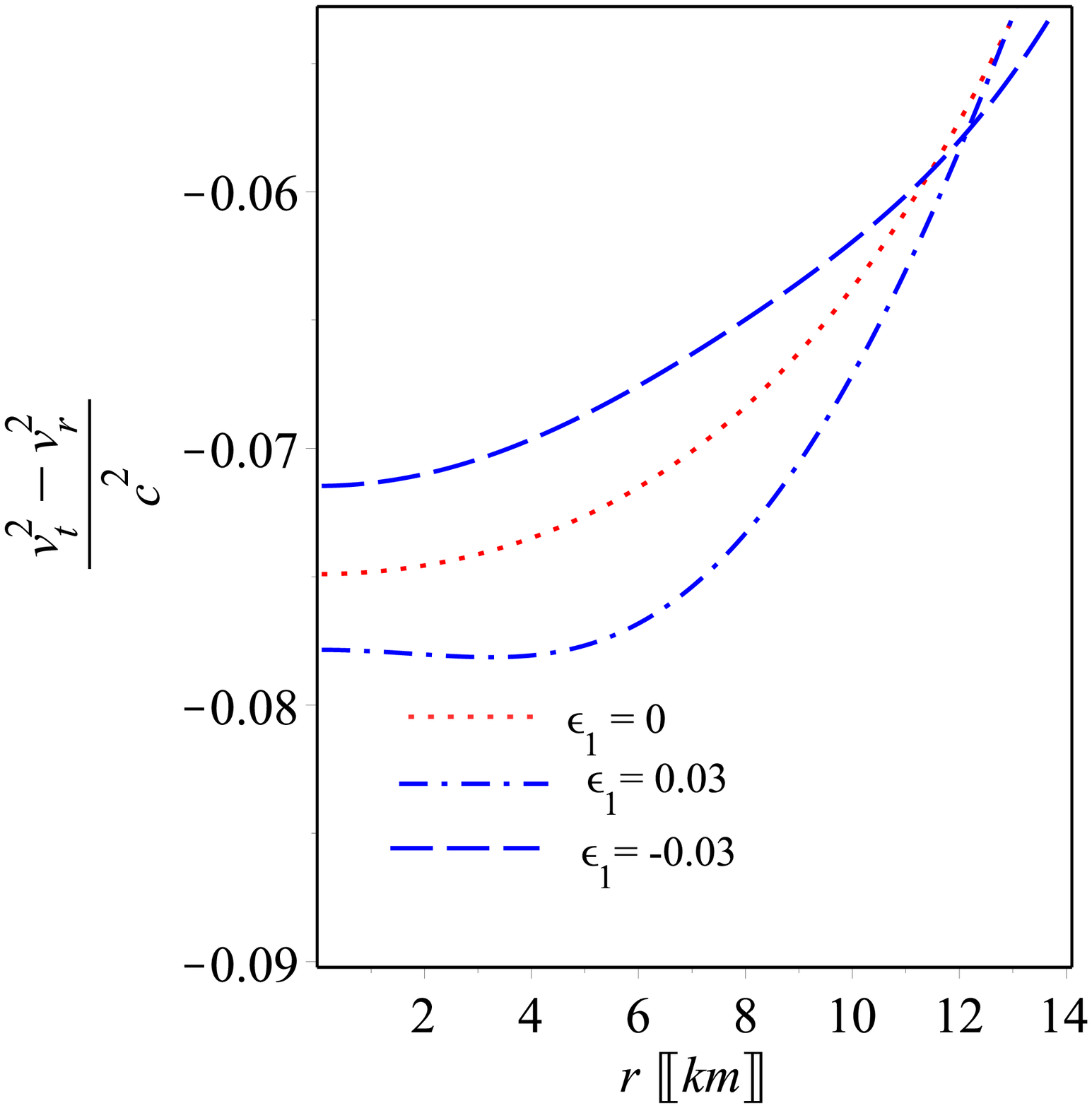}}
\caption{Speed of sound in the fluid inside the pulsar J0740+6620 for $\epsilon_1=0,~ \pm 0.03$: \subref{fig:vr} and \subref{fig:vt} represent the propagation of sound inside the pulsar in radial and tangential directions \eqref{eq:sound_speed}. \subref{fig:vt-vr} the plots ensure that the model verifies the stability constraint $(v_t^2-v_r^2)/c^2 < 0$ of strong anisotropic case.}
\label{Fig:Stability}
\end{figure*}

One of the most important features which characterizes physical structures is the causality condition, that is the speed of sound in a fluid cannot exceed the speed of light. Recalling the induced EoSs \eqref{eq:KB_EoS2}, the radial and tangential speed of sound are the slopes of these linear relations, i.e.
\begin{equation}\label{eq:sound_speed}
  v_r^2 =  \frac{ d{ p}_r}{d { \rho}}=  \frac{p'_r}{{ \rho'}}, \quad
  v_t^2 = \frac{d{  p}_t}{d{   \rho}}= \frac{p'_t}{{ \rho'}}.
\end{equation}
By virtue of Eqs. \eqref{sol}, we obtain the density and pressure gradients as given in appendix \ref{Sec:App_2}, see Eqs. \eqref{eq:dens_grad}--\eqref{eq:pt_grad}. We visualize the sound speed propagation in the radial and the tangential directions inside the pulsar J0740+6620 for different values of the model parameter $\epsilon_1$ as shown in Figs. \ref{Fig:Stability}\subref{fig:vr} and \subref{fig:vt}. The plots ensure that $0\leq {v_r^2}/c^2\leq 1$ and $0\leq {v_t^2/c^2} \leq 1$ which fulfill the stability and causality conditions. In addition, in Fig. \ref{Fig:Stability}\subref{fig:vt-vr}, we show that $-1< (v_t^2-v_r^2)/c^2 < 0$ everywhere inside the pulsar J0740+6620 as required for anisotropic stellar configuration to be stable \citep{Herrera:1992lwz}.

We note that the sound speed in the radial and the tangential directions vary with the radial distance as shown in Figs. \ref{Fig:Stability}\subref{fig:vr} and \subref{fig:vt}. For $\epsilon_1=0.03$, we find $0.30< v_r^2/c^2<0.41$ and $0.17<v_t^2/c^2<0.31$. For $\epsilon_1=-0.03$, we find $0.32< v_r^2/c^2<0.35$ and $0.21<v_t^2/c^2<0.22$. The upper values of these intervals give the sound speed at the center, which are in perfect agreement with the corresponding values previously obtained in Subsection \ref{Sec:matt} from the induced EoSs \eqref{eq:KB_EoS2}, those are ($\epsilon_1=0.03, v_r^2\approx 0.41c^2, v_t^2\approx 0.31c^2$) and ($\epsilon_1=-0.03, v_r^2\approx 0.35c^2, v_t^2\approx 0.22c^2$).
\subsection{Adiabatic indices and hydrodynamic equilibrium}\label{Sec:TOV}
\begin{figure}
\centering
\subfigure[~The adiabatic index]{\label{fig:gamar1}\includegraphics[scale=0.28]{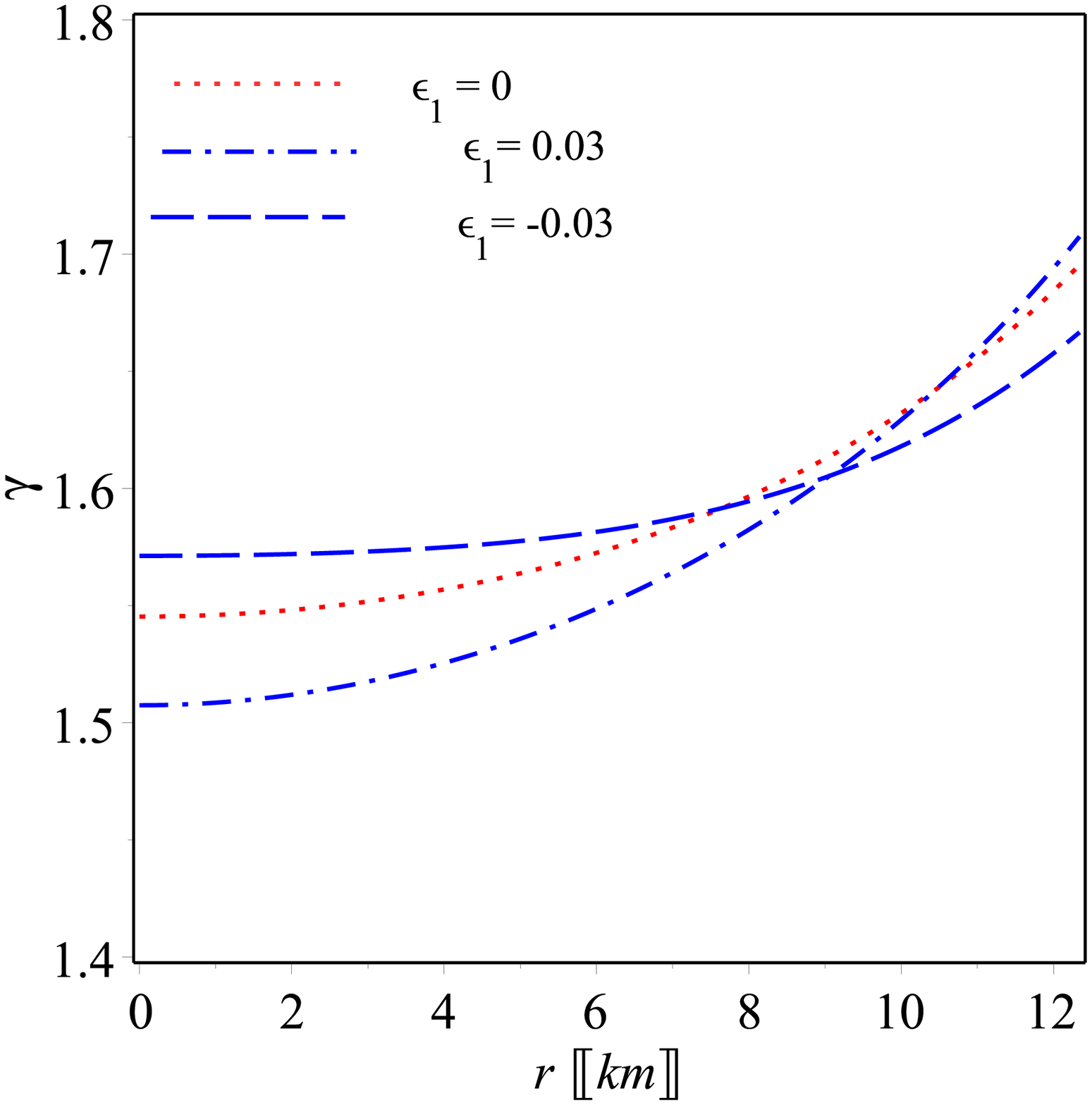}}\hspace{0.2cm}
\subfigure[~Radial adiabatic index]{\label{fig:gamar}\includegraphics[scale=0.28]{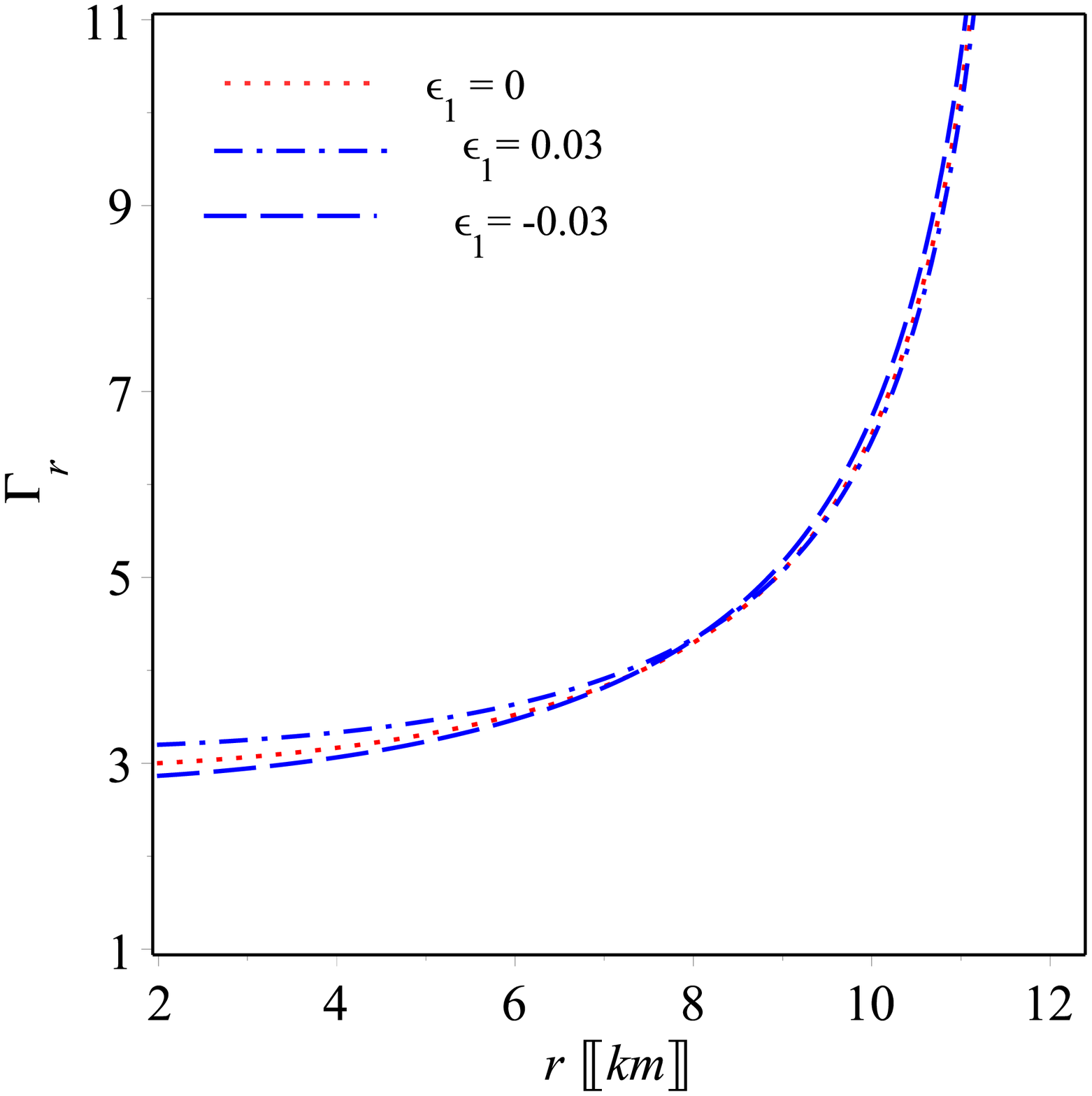}}\hspace{0.2cm}
\subfigure[~Tangential adiabatic index]{\label{fig:gamar}\includegraphics[scale=0.28]{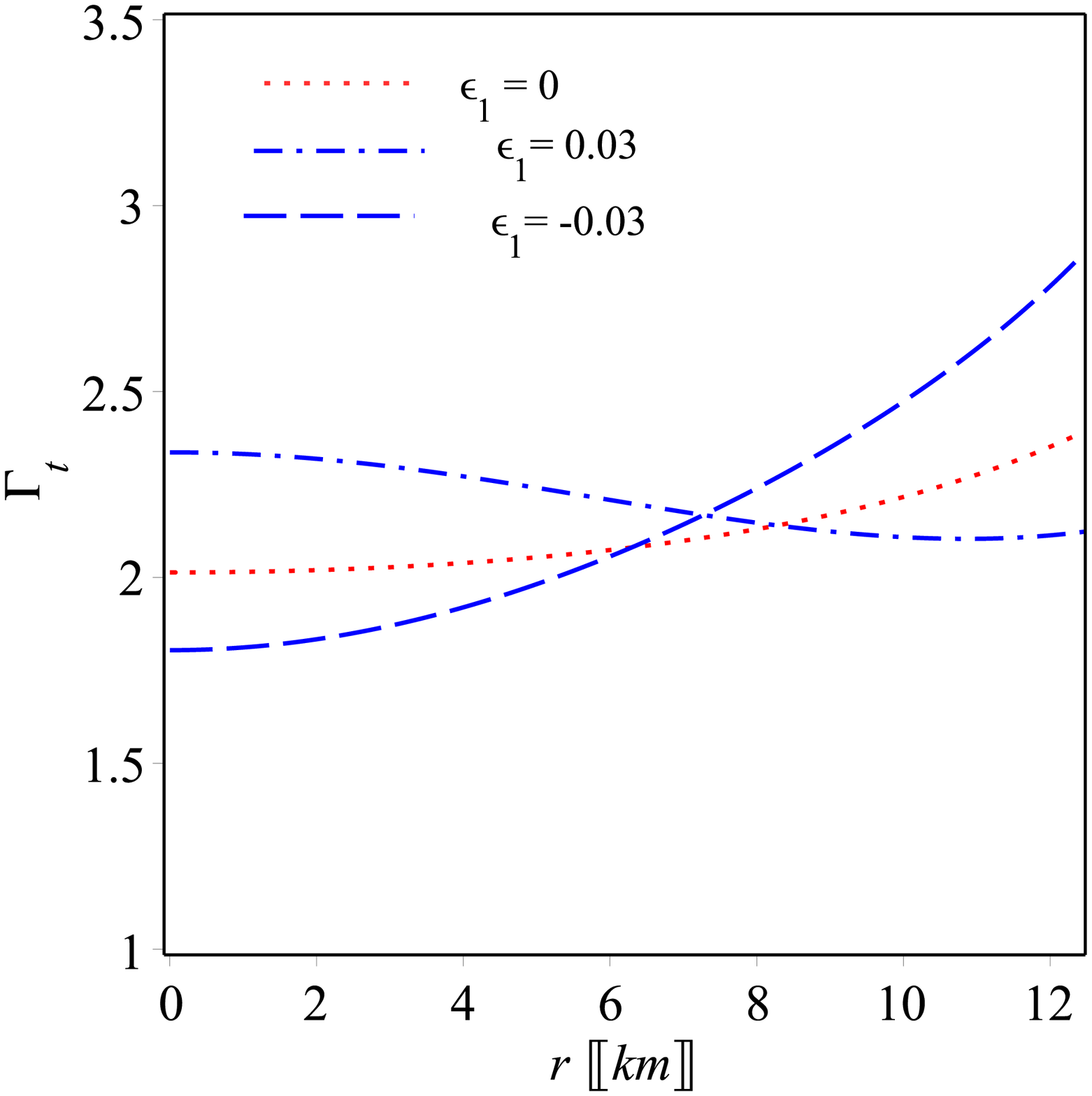}}
\caption{The adiabatic indices given by Eq.~\eqref{eq:adiabatic} of the pulsar J0740+6620. The plots ensure that the pulsar is stable, since $\gamma>4/3$ and $\Gamma_r>\gamma$ and $\Gamma_t>\gamma$ everywhere inside the pulsar as required for strong anisotropic fluid.}
\label{Fig:Adiab}
\end{figure}

It is known that Newtonian gravity cannot put an upper limit on the mass if the adiabatic index (ratio of specific heats) for a given EoS exceeds $4/3$, i.e. $\gamma>4/3$. In other words for a stable configuration in Newton's gravity $\gamma<4/3$. However full relativistic anisotropic neutron star model shows that the star can be stable against radial perturbation where $\gamma>4/3$. We therefore define the adiabatic index \citep{Chandrasekhar:1964zz,chan1993dynamical}
\begin{equation}\label{eq:adiabatic}
{\gamma}=\frac{4}{3}\left(1+\frac{{ \Delta}}{r|{  p}'_r|}\right)_{max},\qquad \qquad
{\Gamma_r}=\frac{{\rho c^2}+{p_r}}{{p_r}}{v_r^2}, \qquad \qquad
{\Gamma_t}=\frac{{\rho c^2}+{p_t}}{{p_t}}{v_t^2 .}
\end{equation}
Clearly for isotropic case, $\Delta=0$, one obtains $\gamma=4/3$, for mild anisotropic case, $\Delta<0$, one obtains $\gamma<4/3$ similar to Newtonian theory, and for strong anisotropic case, $\Delta>0$, similar to the present study, one obtains $\gamma>4/3$. Neutral equilibrium occurs for $\Gamma=\gamma$ and stable equilibrium requires $\Gamma> \gamma$ \cite{chan1993dynamical,1975A&A....38...51H}. Using the field equations \eqref{sol} and the gradients \eqref{eq:pr_grad}--\eqref{eq:pt_grad}, we show that the quadratic gravity theory provides a stable anisotropic model of the pulsar J0740+6620 for both values of the model parameter $\epsilon_1$ as presented by Fig. \ref{Fig:Adiab}.

\begin{figure}
\centering
\subfigure[~TOV constraints ($\epsilon=0$, GR)]{\label{fig:GRTOV}\includegraphics[scale=0.28]{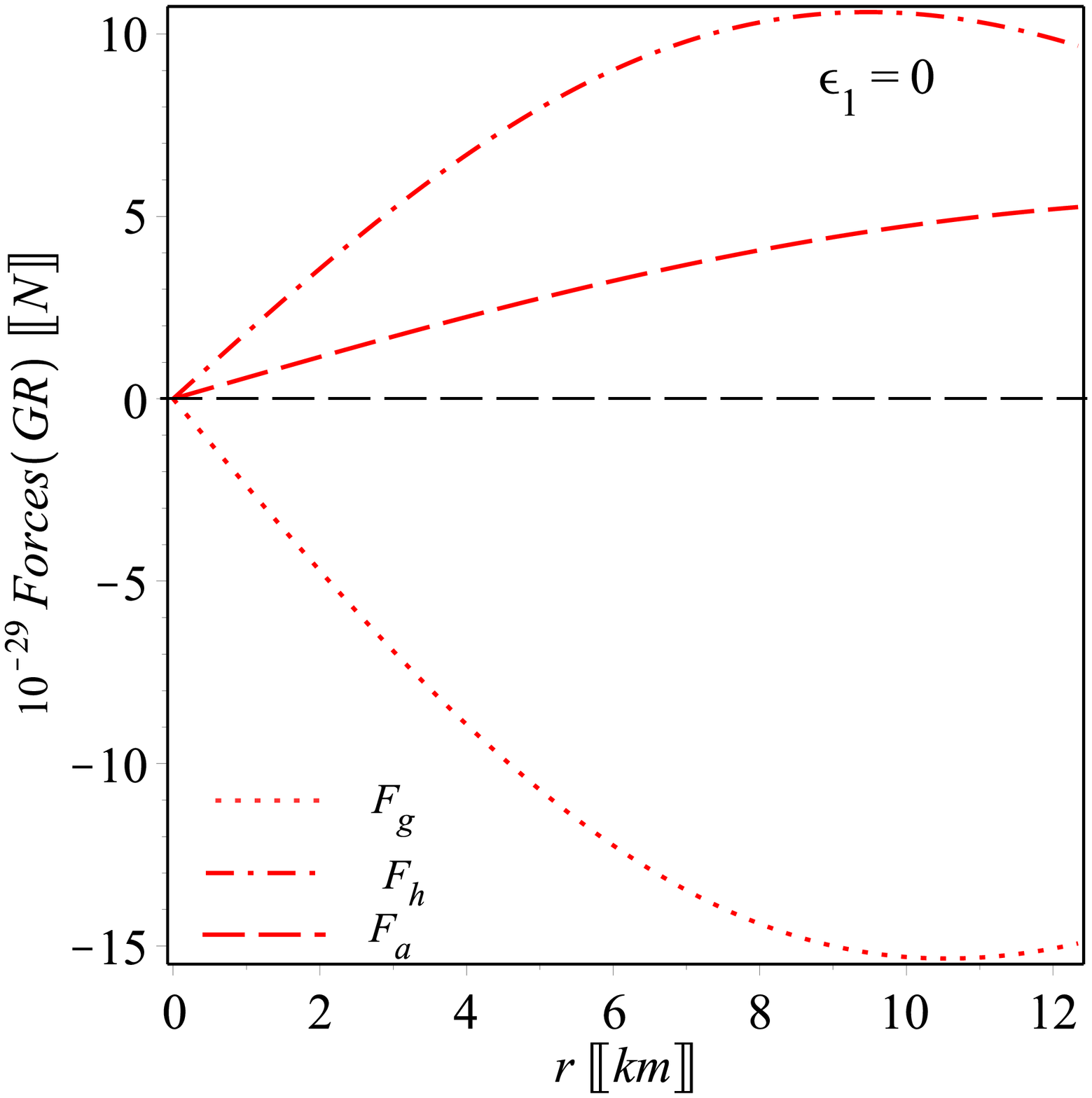}}\hspace{0.2cm}
\subfigure[~TOV constraints ($\epsilon>0$, $f(R)$)]{\label{fig:FRp}\includegraphics[scale=0.28]{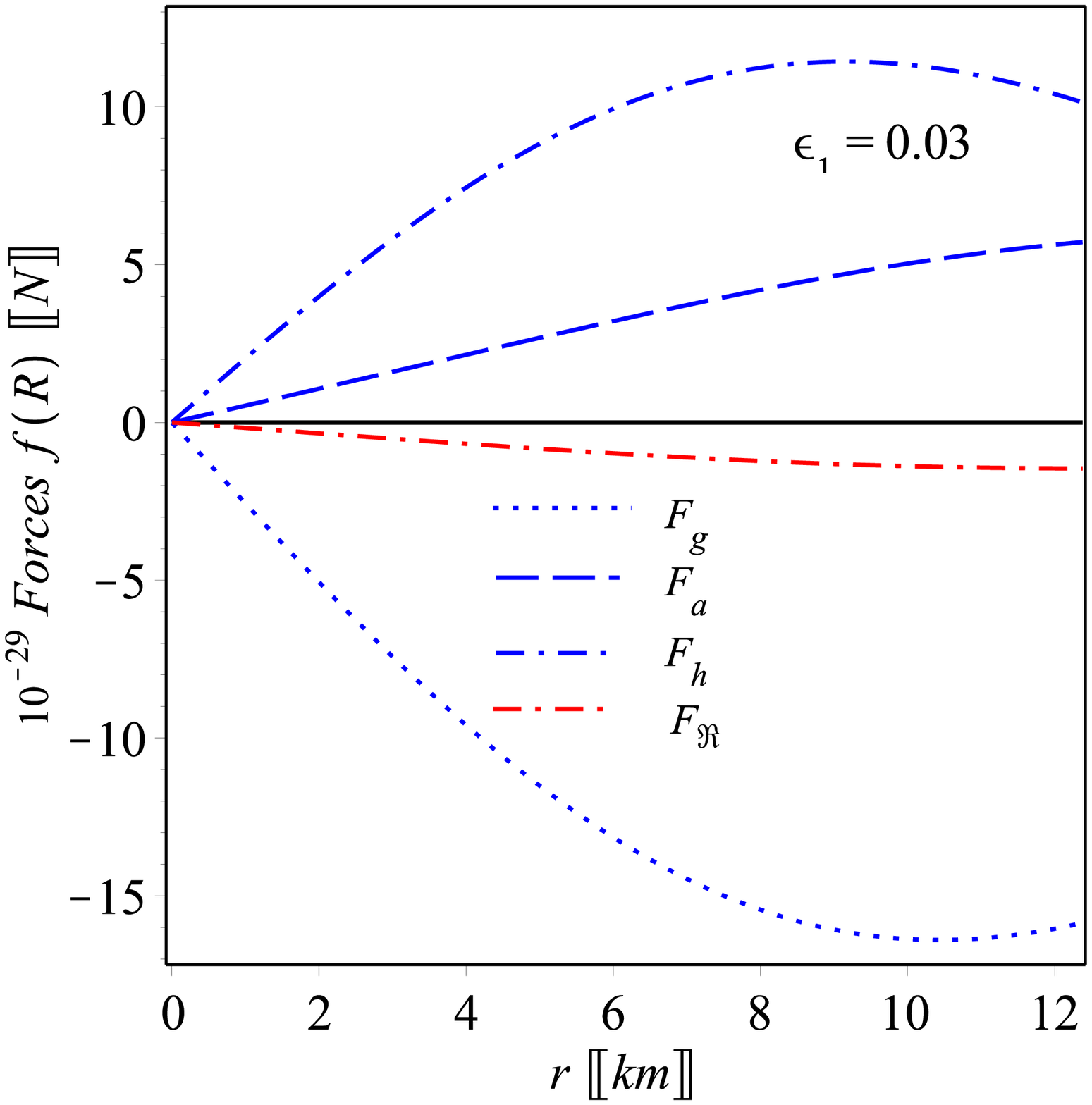}}\hspace{0.2cm}
\subfigure[~TOV constraints ($\epsilon<0$, $f(R)$)]{\label{fig:FRn}\includegraphics[scale=0.28]{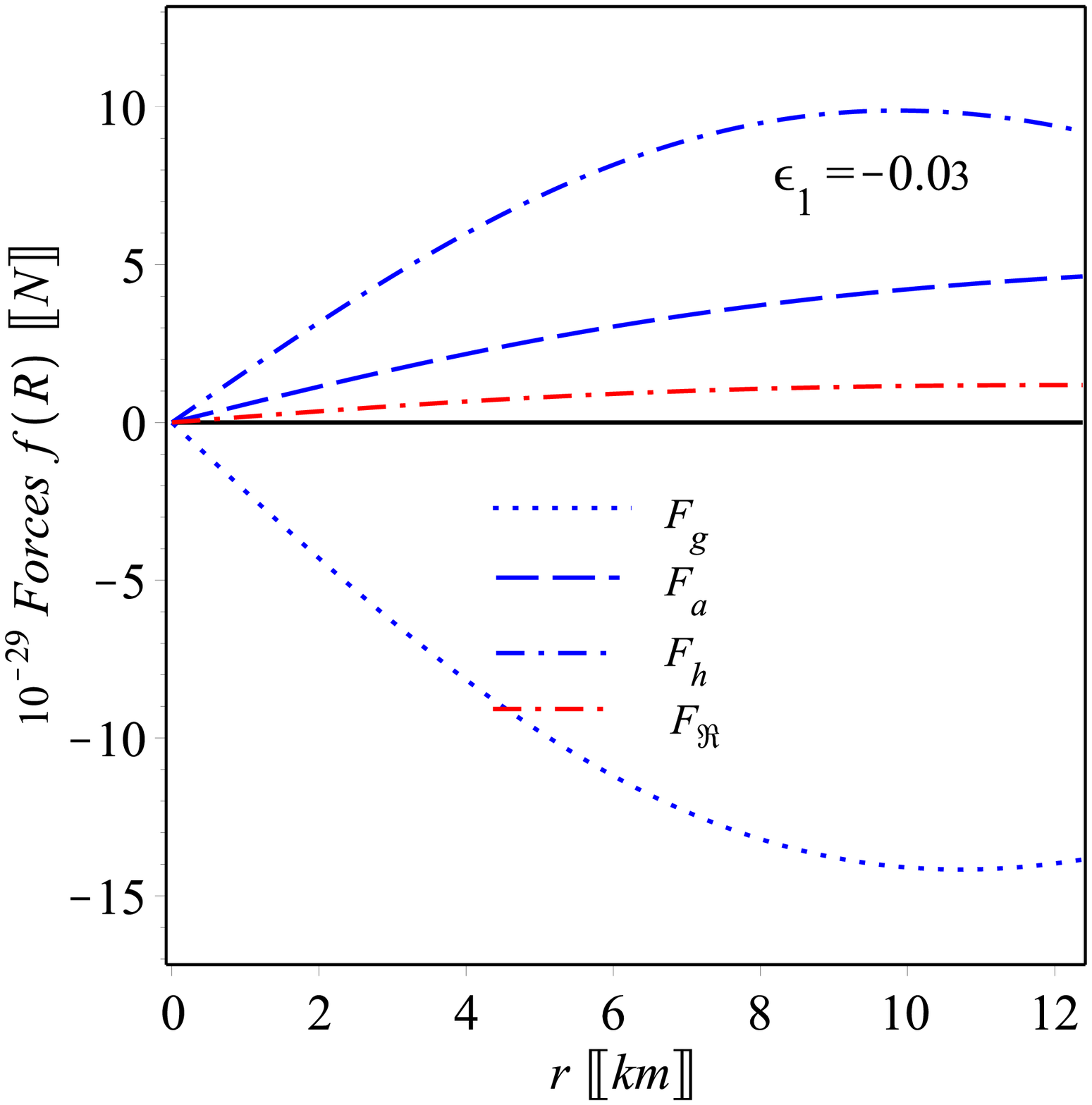}}
\caption{Tolman–Oppenheimer–Volkoff constrains \eqref{eq:RS_TOV}: Different forces, as obtained by Eqs. \eqref{eq:Forces}, inside the pulsar J0740+6620 are represented for $\epsilon_1=0,~\pm 0.03$. For $\epsilon_1=0.03$ the quadratic correction contributes by an additional negative force which strengthens the gravitational collapse force. For $\epsilon_1=-0.03$ the quadratic correction contributes by an additional positive force which weakens the gravitational collapse force.}
\label{Fig:TOV}
\end{figure}

We next investigate the hydrodynamical equilibrium of the present model via the TOV equation, which has been modified for a given $f(R)$ theory as follows
\begin{equation}\label{eq:RS_TOV}
{\mathit F_a}+{\mathit F_g}+{\mathit F_h}+{\mathit F_R=0}\,,
\end{equation}
where $F_a$, $F_g$, and $F_h$ are the usual anisotropic, gravitational, and hydrostatic forces in addition to the extra force $F_R$ due to the counterpart of the $f(R)$ theory (i.e. quadratic gravity $\epsilon R^2$). These are defined as
\begin{eqnarray}\label{eq:Forces}
  {\mathit F_a} =&\frac{ 2{\mathit  \Delta}}{\mathit r} ,\qquad
  {\mathit F_g} = -\frac{{\mathit  M_g}}{r}({\mathit  \rho c^2}+{\mathit p_r})e^{{\mathit  \gamma/2}} ,\qquad\nonumber\\
  {  F_h} =&-{\mathit  p'_t} ,\qquad
  { F_R} = 2\epsilon_1({  c^2 \rho}'-{ p}'_r-2{  p}'_{t}).
\end{eqnarray}
In $F_g$ force equation, we introduced $\delta:=\psi-\lambda$, and the quantity $M_g$ denotes the gravitational of the an isolated system in 3-spaces ${\mathit V}$($t=$ constant), which can be defined by Tolman mass formula \citep{1930PhRv...35..896T} in $f(R)$ gravity
\begin{eqnarray}\label{eq:grav_mass}
{\mathit M_g(r)}&=&{\int_{\mathit V}}\Big(\mathbb{{\bar{\mathfrak T}}}{^r}{_r}+\mathbb{\bar{\mathfrak T}}{^\theta}{_\theta}+\bar{\mathfrak{T}}{^\phi}{_\phi}-\bar{\mathfrak{T}}{^t}{_t}\Big)\sqrt{-g}\,dV\nonumber\\
&=&e^{-\psi}(e^{\psi/2})'  e^{\lambda/2} r =\frac{1}{2} r \psi' e^{-\delta/2},
\end{eqnarray}
and therefore the gravitational force reads ${\mathit F_g}=-\frac{a_0 r}{R_s^2}({\mathit \rho c^2}+{ p_r})$. Using the field equations \eqref{sol} and the gradients \eqref{eq:pr_grad}--\eqref{eq:pt_grad}, we show that the quadratic gravity theory satisfies \eqref{eq:RS_TOV} providing a stable model of the pulsar J0740+6620 for both values of the model parameter $\epsilon_1$ including the GR case ($\epsilon_1=0$) as presented by Fig. \ref{Fig:TOV}.

We note that the strong anisotropy condition, $\Delta>0$, induces a positive force against the negative gravitational (collapsing) force. This plays an important role to increase the size of the star allowing for more mass (compactness) while keeping a stable configuration. Recognizably, in Figs. \ref{Fig:TOV}\subref{fig:FRp} and \subref{fig:FRn}, the extra force due to quadratic gravity contributes to support the gravitational collapse if $\epsilon$ is positive, and it contributes to partly oppose gravitational collapse if $\epsilon$ is negative. This analysis is confirmed by the results previously obtained for the pulsar J0740+6620 in Subsection \ref{Sec:obs_const}, those are ($\epsilon_1=0.03, M=2.174 M_\odot, {R_s} =12.07~\text{km}, C=0.56$) and ($\epsilon_1=-0.03, M=1.953 M_\odot, {R_s} =12.88~\text{km}, C=0.49557$).
\section{Equation of stat and Mass-radius relation}\label{Sec:EoS_MR}

\begin{figure}[th!]
\centering
\subfigure[~Radial EoS]{\label{fig:RfEoSp}\includegraphics[scale=0.45]{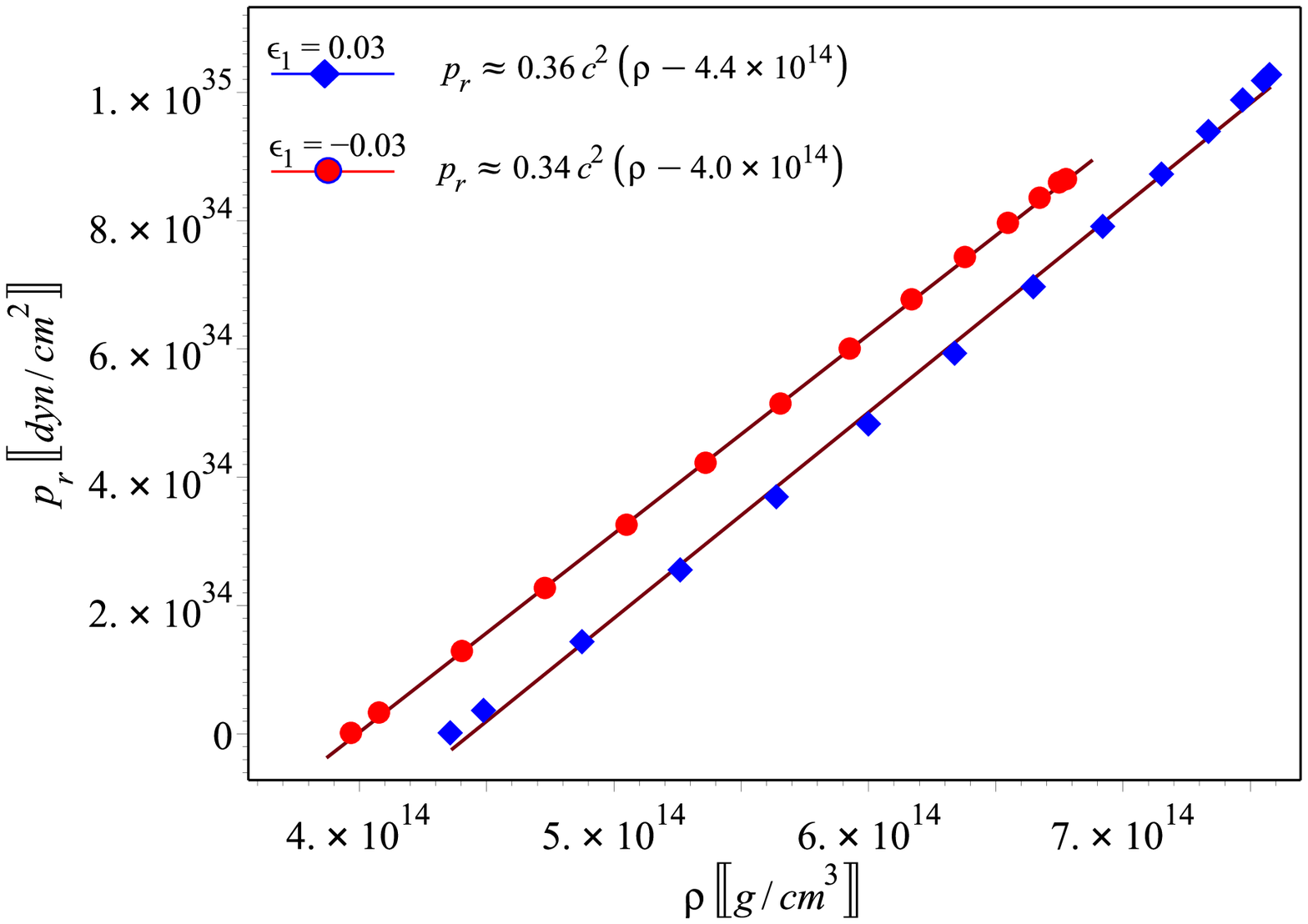}}
\subfigure[~Tangential EoS]{\label{fig:TEoSn}\includegraphics[scale=0.45]{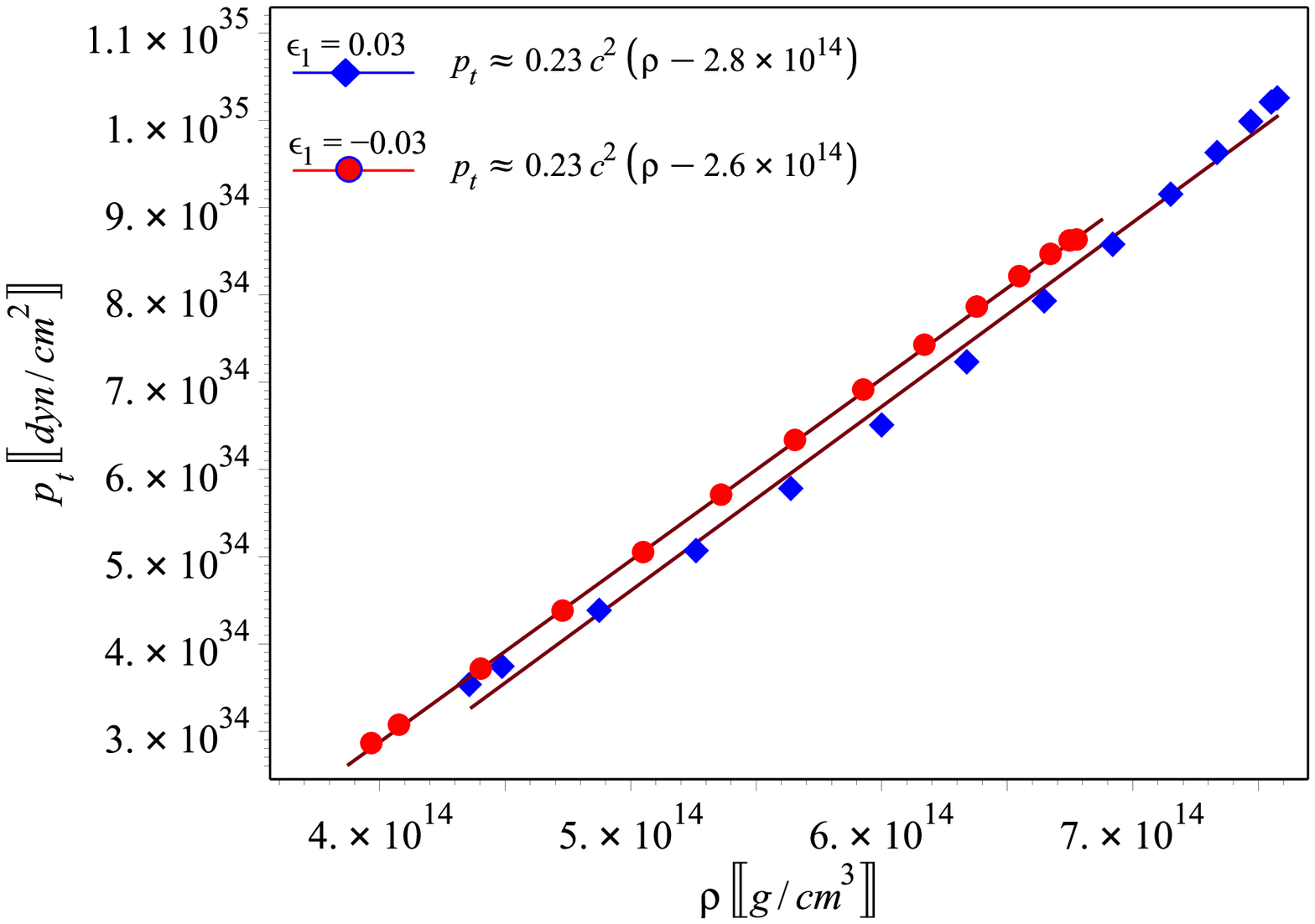}}
\caption{The best fit EoSs of the pulsar J0740+6620: \subref{fig:RfEoSp} We generate a sequence of points for the density and radial pressure by virtue of Eqs. \eqref{sol} for $\epsilon=\pm 0.03$, the points fit well with linear EoS pattern. \subref{fig:TEoSn} Similarly for tangential EoSs with $\epsilon=\pm 0.03$ the points fit well with linear model. The best fit linear EoSs are in agreement with the previously obtained ones, namely \eqref{eq:KB_EoS2}, which verifies the validity of those relations everywhere inside the pulsar. Notably slight deviations from linear pattern can be observed in the case where $\epsilon_1=0.03$, which indicates that quadratic polynomial could fit better, i.e. $p_{r,t}(\rho)\approx \tilde{c}_0+\tilde{c}_1\rho+\tilde{c}_2\rho^2$, in this case.}
\label{Fig:EoS}
\end{figure}

The nature of matter inside neutron star cores is still puzzling many astrophysicists since their core densities reach a few times the nuclear saturation density which is not accessible by terrestrial laboratories. Although the EoS of the neutron star matter is unknown, astrophysical observations of neutron stars mass and radius could constrain it or at least exclude some. In this sense, the mass-radius diagram for a given EoS could be constrained by astrophysical observations. Indeed, we do not impose EoSs in this study, we instead use KB ansatz \eqref{eq:KB}. However, this ansatz relates the pressures and the density as shown by the induced EoSs \eqref{eq:KB_EoS2}, which are mostly valid at the center due to power series assumptions. We confirm the validity of those equations by generating a sequence of density and pressure values from the center to the surface for positive and negative values of the model parameter $\epsilon_1$. Using the numerical values given in \ref{Sec:obs_const} for the pulsar J0740+6620 and the field equations of the quadratic $f(R)$ gravity, namely Eqs. \eqref{sol}, we generate the sequences as presented in Fig. \ref{Fig:EoS}.

Clearly, the data fits well with a linear model in both cases: For $\epsilon_1=0.03$, the best-fit equations can be written as $p_r \text{[dyn/cm$^2$]}\approx 0.36 c^2(\rho-4.4 \times 10^{14}\text{[g/cm$^3$]})$ and $p_t \text{[dyn/cm$^2$]}\approx 0.23c^2(\rho-2.8\times 10^{14}\text{[g/cm$^3$]})$. For $\epsilon_1=-0.03$, the best fit equations can be written as $p_r \text{[dyn/cm$^2$]}\approx 0.34 c^2(\rho-4.0 \times 10^{14}\text{[g/cm$^3$]})$ and $p_t \text{[dyn/cm$^2$]}\approx 0.23c^2(\rho-2.6\times 10^{14}\text{[g/cm$^3$]})$. Those are in agreement with the previously obtained EoSs \eqref{eq:KB_EoS2}, explicitly $p_r \text{[dyn/cm$^2$]}\approx 0.41 c^2(\rho-4.82 \times 10^{14}\text{[g/cm$^3$]})$ and $p_t \text{[dyn/cm$^2$]}\approx 0.31c^2(\rho-3.85\times 10^{14}\text{[g/cm$^3$]})$ for $\epsilon_1=0.03$, and $p_r \text{[dyn/cm$^2$]}\approx 0.35 c^2(\rho-4.0 \times 10^{14}\text{[g/cm$^3$]})$ and $p_t \text{[dyn/cm$^2$]}\approx 0.22c^2(\rho-2.49\times 10^{14}\text{[g/cm$^3$]})$ for $\epsilon_1=-0.03$. We note that the linear best fit in the positive $\epsilon_1$ case slightly changes the sound speeds and the surface density, unlike the negative case which perfectly coincides with the induced EoSs. In this sense, we find that quadratic polynomial, i.e. $p_{r,t}(\rho)\approx \tilde{c}_0+\tilde{c}_1\rho+\tilde{c}_2\rho^2$, could provide better fitting in the positive $\epsilon_1$ case, those are non-linear EoSs.
\begin{figure*}[t]
\subfigure[~Compactness-radius diagram]{\label{fig:Comp}\includegraphics[scale=0.4]{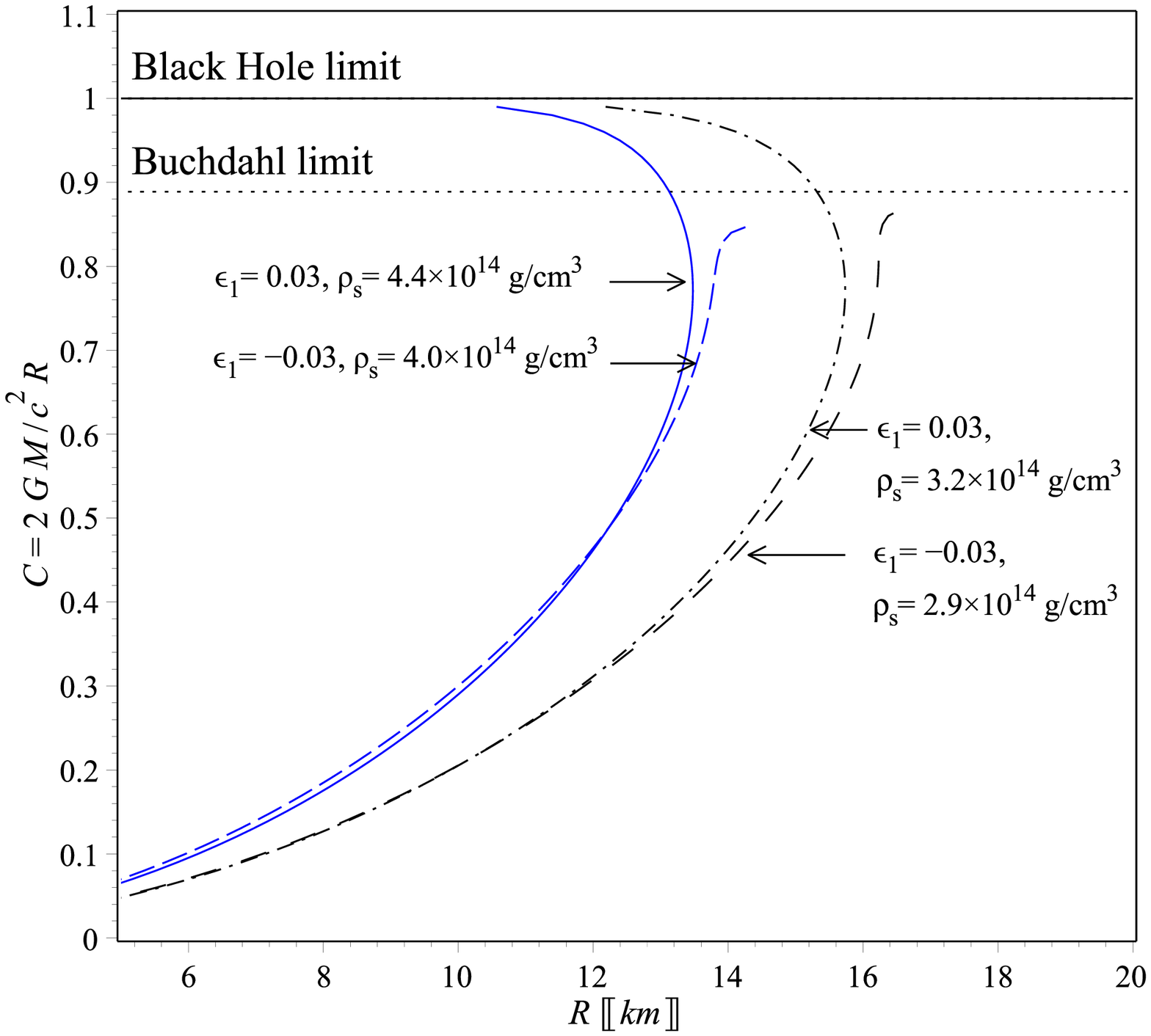}}\hspace{0.5cm}
\subfigure[~Mass-radius diagram]{\label{fig:MR}\includegraphics[scale=.4]{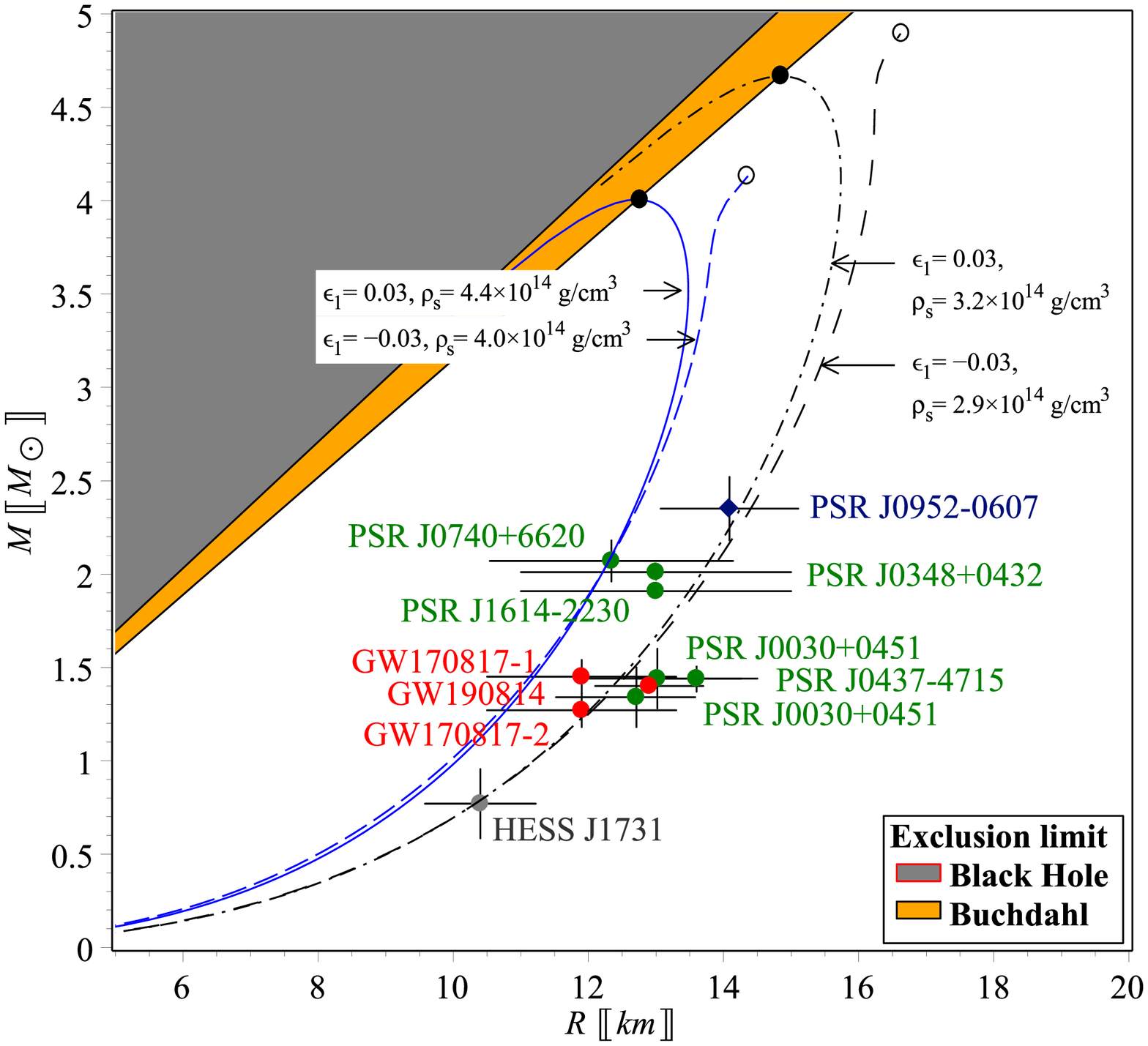}}
\caption{\subref{fig:Comp} Compactness-radius (CR) diagram: The solid (dotted) horizontal line represents the BH (Buchdahl) limit on the compactness $C=1$ ($C=8/9$). We plot the CR curves corresponds to the best fit EoSs as given in Fig. \ref{Fig:EoS}. In addition we plot CR curves for different boundary densities $\rho_s=3.2\times 10^{14}$ g/cm$^3$ and $\rho_s=2.9\times 10^{14}$ g/cm$^3$. We note that for $\epsilon_1=0.03$ the compactness is extendable to the BH limit $C\to 1$, this feature is common with GR model \citep{Roupas:2020mvs}. Interestingly for $\epsilon_1=-0.03$ the maximum compactness is below Buchdal upper bound which differentiates the quadratic gravity theory from GR in this case. \subref{fig:MR}  Mass-radius (MR) diagram: The gray (orange) region represents the BH (Buchdahl) limit. Similarly, for positive $\epsilon$, the MR curves are extendable to nonphysical branches exceeding Buchdahl limit with maximum mass at this limit as represented by solid circles, while for negative $\epsilon$ the MR curves do not cross Buchdahl limit with maximum mass as represented by open circles.}
\label{Fig:CompMR}
\end{figure*}

One of the important constraint on stable stellar configuration has been derived by Buchdahl in terms of an upper bound on the compactness value \cite{PhysRev.116.1027}, that is $C<8/9$. In fact this bound has been derived for an isotropic (or mild anisotropic) spherically symmetric GR solution in particular. However, it has been shown that this limit can be violated by dropping one or more of these assumptions. In the more realistic case of strong anisotropic models, the compactness can arbitrarily approaches the black hole limit, i.e. $C\to 1$, even in GR \cite{Alho:2022bki}. Remarkably, other physical constraints put more restrictive bounds on the compactness in strong anisotropic cases \cite{Alho:2021sli,Roupas:2020mvs,Raposo:2018rjn,Cardoso:2019rvt}. Same conclusion has been obtained when nonminimal coupling between geometry and matter is considered \citep{Nashed:2022zyi,ElHanafy:2022kjl,2023arXiv230514953E}. In this sense, it is useful to examine this constraint within the quadratic $f(R)$ gravity as considered in the present study. For a generalized $f(R)$ gravity, Buchdahl limit is given by \cite{Goswami:2015dma}
\begin{align}
    C=\frac{2GM}{c^2 R_s}< \frac{4\frac{f_R(R_s)}{f_R(0)}\left[1+\frac{f_R(R_s)}{f_R(0)}\right]}{\left[1+2\frac{f_R(R_s)}{f_R(0)}\right]^2}.\nonumber
\end{align}
For the present model of quadratic $f(R)$ gravity, in particular, we write the modified Buchdahl upper limit on the compactness
\begin{equation}
    C < {{4(\frac{1}{2}+\epsilon R|_{r=R_s})\left[1+\epsilon(R|_{r=0}+R|_{r=R_s})\right]} \over {\left[\frac{3}{2}+\epsilon(R|_{r=0}+2 R|_{r=R_s})\right]^2}}.
\end{equation}
We remind that $\epsilon=\epsilon_1 \ell^2$ with $\ell$ being the radius of canonical neutron star $10$ km. Obviously the above inequality reduces to Buchdahl's limit in GR if $\epsilon=0$, i.e. $C<8/9$ \cite{PhysRev.116.1027}.

In the present study of the pulsar J0740+662, we calculate the Buchdahl limit: For $\epsilon_1=0.03$ we obtain $C\lesssim 0.888$, and for $\epsilon_1=-0.03$ we find $C\lesssim 0.891$ which slightly modifies the GR Buchdahl limit. Notably the negative $\epsilon_1$ values increase the upper limit on the compactness. This is in agreement with our conclusion that the quadratic gravity in this case contributes to the hydrodynamic equilibrium (TOV) equation by an additional force that opposes the gravitational force and allows the star to contain more mass and higher compactness values. Since both cases slightly change Buchdahl limit, we use the standard constraint, $C\leq 8/9$, as represented by the horizontal dotted line in the compactness-radius diagram in Fig. \ref{Fig:CompMR}\subref{fig:Comp}. Firstly, we take the following pairs ($\epsilon_1=0.03$, $\rho_\text{s}=4.4\times 10^{14}$ g/cm$^3$) and ($\epsilon_1=-0.03$, $\rho_\text{s}=4.0\times 10^{14}$ g/cm$^3$) at the surface as obtained by the best-fit EoSs, for arbitrary values of the compactness parameter $0 \leq C \leq 1$, we solve the density profile \eqref{sol} for the radius $R$. Similarly, we obtain a compactness-radius curves corresponds to the boundary conditions ($\epsilon_1=0.03$, $\rho_\text{s}=3.2\times 10^{14}$ g/cm$^3$) and ($\epsilon_1=-0.03$, $\rho_\text{s}=2.9\times 10^{14}$ g/cm$^3$) as presented in Fig. \ref{Fig:CompMR}\subref{fig:Comp}. Although the maximum compactness approaches the black hole limit $C\to 1$ when $\epsilon_1$ is positive similar to GR \citep{Roupas:2020mvs}, interestingly for negative value of $\epsilon_1$ the maximum compactness $C_\text{max}=0.85$ ($C_\text{max}=0.87$), where $\rho_\text{s}=4.0\times 10^{14}$ g/cm$^3$ ($\rho_\text{s}=2.9\times 10^{14}$ g/cm$^3$), is below Buchdahl upper bound on the compactness.
\begin{table}[t]
    \centering
    \caption{Astrophysical observations of mass and radius of some pulsars.}
    \begin{tabular}{cccc}
    \hline
       Pulsar                  & Mass ($M_\odot$)       & Radius (km)             & Ref.\\[5pt]
\hline
       J0952\textendash{0607}  & $2.35\pm 0.17$         & $14.087 \pm 1.0186$     & \cite{Romani:2022jhd,2023arXiv230514953E}\\[5pt]
       J0348+0432              & $2.01 \pm 0.04$        & $13 \pm 2$              & \citep{Antoniadis:2013pzd}\\[5pt]
       J1614\textendash{2230}  & $1.908 \pm 0.016$      & $13 \pm 2$              & \citep{Demorest:2010bx,Fonseca:2016tux,NANOGRAV:2018hou}\\[5pt]
       J0030+0451              & $1.44^{+0.15}_{-0.14}$ & $13.02_{-1.06}^{+1.24}$ & \citep{Miller:2019cac} \\[5pt]
                               & $1.34^{+0.15}_{-0.16}$ & $12.71^{+1.14}_{-1.19}$ & \citep{Raaijmakers:2019qny}\\[5pt]
       J0437\textendash{4715}  & $1.44 \pm 0.07$        & $13.6 \pm 0.9$          & \citep{Reardon:2015kba,Gonzalez-Caniulef:2019wzi}\\[5pt]
       GW170817-1              & $1.45 \pm 0.09$        & $11.9 \pm 1.4$          & \citep{LIGOScientific:2018cki}\\[5pt]
       GW170817-2              & $1.27 \pm 0.09$        & $11.9 \pm 1.4$          & \citep{LIGOScientific:2018cki}\\[5pt]
       LIGO/Virgo              & $1.4$                  & $12.9\pm 0.8$           & \citep{LIGOScientific:2020zkf}\\[5pt]
       J1731\textendash{347}   & $0.77_{-0.17}^{+0.20}$ & $10.4^{+0.86}_{-0.78}$  & \citep{2022NatAs...6.1444D}\\[5pt]
         \hline
    \end{tabular}
        \label{tab:obs_const}
\end{table}

For the best fit EoS as obtained earlier in this section, we give the corresponding Mass-Radius curves in Fig. \ref{Fig:CompMR}\subref{fig:MR} for both $\epsilon_1$ cases where the gravitational mass is determined by virtue of the matching condition \eqref{eq:bo}, i.e. $M= \frac{c^2 R}{2 G} (1-e^{-a_2})$. Therefore, we take the boundary density $\rho_s=4.4\times 10^{14}$ g/cm$^{3}$ with $\epsilon_1=0.03$, which gives a maximum mass $M=4.0 M_\odot$ at radius $R=12.6$ km. For a boundary density $\rho_s=4.0\times 10^{14}$ g/cm$^{3}$ with $\epsilon_1=-0.03$, it gives a maximum mass $M=4.1 M_\odot$ at radius $R=14.4$ km. Those are in perfect agreement with the pulsar J0740+6620. In addition we use different values of boundary density which still in agreement with the pulsar J0740+6620, but give better fit with other pulsars (see Table \ref{tab:obs_const}) as well. For the boundary density $\rho_s=3.2\times 10^{14}$ g/cm$^{3}$ with $\epsilon_1=0.03$, it gives a maximum mass $M=4.7 M_\odot$ at radius $R=14.6$ km. For $\rho_s=2.9\times 10^{14}$ g/cm$^{3}$ with $\epsilon_1=-0.03$, the maximum mass is $M=4.9 M_\odot$ at radius $R=16.6$ km. Obviously, for the positive $\epsilon_1$ cases, the mass-radius curves are extendable to the black hole limit as represented by the gray region on Fig. \ref{Fig:CompMR}\subref{fig:MR}. However, the maximum masses (solid circles) in that case are almost at Buchdahl limit as represented by the orange region. On the contrary, for negative $\epsilon_1$ value, the mass-radius curve is considerably below Buchdahl limit with slightly higher values of maximum masses (open circles) than the positive $\epsilon_1$ case, but with much larger sizes. Finally, we note that in the case when the surface density near to the nuclear saturation density, it gives a better fit with most of the pulsars. However, this will be in conflict with the best fit EoSs given in Fig. \ref{Fig:EoS}, since $\epsilon_1$ and $\rho_s$ are correlated.

\section{Conclusion}\label{Sec:Conclusion}

In the present study, we investigated the quadratic $f(R)=R+\epsilon R^2$ gravity by confronting the theory with astrophysical observations of pulsars. We assumed the more realistic case of anisotropic fluid as expected for high dense matter inside pulsars. We additionally assumed the interior of the static spherically symmetric stellar model is governed by KB ansatz which guarantee the regularity of the interior spacetime. In particular, we used the accurate measurements of the mass and radius of the PSR J0740+6620, $M=2.07 \pm 0.11 M_\odot$ and radius $R=12.34^{+1.89}_{-1.67}$ km, as inferred by NICER+XMM observations \citep{Legred:2021hdx} to constraint the parameter space of the{\color{red} present model} of quadratic $f(R)$ gravity in the {\color{red}present model}, i.e. \{$\epsilon$, $C$\}. On the other hand, the PSR J0740+6620 is one of the heaviest pulsars which makes it a perfect choice to test modified gravity.

In principal, strong anisotropy induces an extra repulsive force in the TOV equation which increases the size of the star and consequently allows for stable stars with higher compactness and higher maximum mass values relative to the isotropic case. Therefore strong anisotropic star could provide a better framework to deal with heavy compact stars $M\sim 2M_\odot$. On the other hand, for the heavy compact stars, it lowers the radial sound speed in comparison to the isotropic case and consequently predicts softer EoS in agreement with the deformabilities of the observed gravitational wave signals.

We showed that the viable range of the quadratic gravity parameter is $|\epsilon_1| < 0.03$ (i.e. $|\epsilon| < 3$ km$^2$). We confirmed the viability of the obtained model via several stability conditions on geometrical and matter sectors. The maximum density at the center of the pulsar J0740+6620 are predicted as follows: For $\epsilon=0.03$, the core density ${\rho_\text{core}\approx 7.58\times 10^{14}}$ g/cm$^{3} \approx 2.8\rho_\text{nuc}$. For $\epsilon=-0.03$, the core density ${\rho_\text{core}\approx 6.78\times 10^{14}}$ g/cm$^{3} \approx 2.5\rho_\text{nuc}$. At the surface, for $\epsilon_1=0.03$, we obtained ${\rho_s\approx 4.36\times 10^{14}}$ g/cm$^{3} \approx 1.6 \rho_\text{nuc}$. For $\epsilon=-0.03$, we found ${\rho_s\approx 3.97\times 10^{14}}$ g/cm$^{3} \approx 1.4\rho_\text{nuc}$. Although we did not impose EoSs in this study, {\color{red}We} proved that the KB model, up to $O(\eta^4)$ where $\eta:=r/R_s$, can relate the pressures and density inside the pulsar by linear relations $p_r(\rho)\approx v_r^2(\rho-\rho_{1})$ and $p_t(\rho) \approx v_t^2 (\rho-\rho_{2})$. More precisely, {\color{red}for for} $\epsilon_1=0.03$, we estimated the EoSs $p_r\approx 0.41 c^2(\rho-4.82 \times 10^{14})$ and $p_t\approx 0.31c^2(\rho-3.85\times 10^{14})$. For $\epsilon_1=-0.03$, we obtained $p_r\approx 0.35 c^2(\rho-4.0 \times 10^{14})$ and $p_t\approx 0.22c^2(\rho-2.49\times 10^{14})$. On the other hand, we generated sequences of exact values of the pressures and density which are in a perfect agreement with linear EoS patterns. For $\epsilon_1=0.03$, the best-fit equations can be written as $p_r\approx 0.36 c^2(\rho-4.4 \times 10^{14})$ and $p_t\approx 0.23c^2(\rho-2.8\times 10^{14})$. For $\epsilon_1=-0.03$, the best fit equations can be written as $p_r\approx 0.34 c^2(\rho-4.0 \times 10^{14})$ and $p_t \approx 0.23c^2(\rho-2.6\times 10^{14})$. This proves the validity of the obtained EoSs everywhere inside the pulsar. In particular for positive $\epsilon$ we observed slight deviations from the linear pattern to quadratic polynomial one, i.e. $p_{r,t}(\rho)\approx \tilde{c}_0+\tilde{c}_1\rho+\tilde{c}_2\rho^2$.

We calculated the modified Buchdahl limit on the maximum compactness corresponds to the quadratic $f(R)$ counterpart which have been obtained to be $C\approx 0.888$ for $\epsilon_1=0.03$ and $C\approx 0.891$ for $\epsilon_1=-0.03$ with slight deviation from the corresponding GR value $C=8/9$. We noted that the present model of strong anisotropic fluid, {\color{red}within in} quadratic gravity framework, cannot put an upper limit on the compactness, in the case of positive $\epsilon_1$, where it can arbitrarily approach the BH limit $C\to 1$. This is a common feature in the models which describe strong anisotropic fluids even in GR. Interestingly, for quadratic gravity with negative $\epsilon_1$ the model restricts the maximum allowed compactness to values below Buchdahl limit. We related this result to the additional repulsive force induced by the quadratic gravity when $\epsilon_1$ is negative, see Fig. \ref{Fig:TOV}\subref{fig:FRn}. This force allows the pulsar to gain more size while its mass is almost fixed at highest values as indicated by Fig. \ref{Fig:CompMR}\subref{fig:Comp}. In the positive $\epsilon$ case, this induced force support gravitational collapse just as in the GR gravity but stronger. This may favor the quadratic gravity with negative $\epsilon_1$ value over the GR scenario.

For the best-fit EoSs we plotted the corresponding mass-radius diagram in Fig. \ref{Fig:CompMR}\subref{fig:MR}. For $\epsilon_1=0.03$ with a boundary density $\rho_s=4.4\times 10^{14}$ g/cm$^{3}$ we got a maximum mass $M=4.0 M_\odot$ at radius $R=12.6$ km. For $\epsilon_1=-0.03$ with a boundary density $\rho_s=4.0\times 10^{14}$ g/cm$^{3}$, we got a maximum mass $M=4.1 M_\odot$ at radius $R=14.4$ km. Those are in perfect agreement with the pulsar J0740+6620. For better fit with other observations from NICER and LIGO, we manged to change the boundary densities to lower values which is in conflict with the predicted values from the best-fit equations of state. Also, as noted that the strong anisotropy and the quadratic gravity (with negative $\epsilon$) contribute to oppose gravitational collapse and effectively lower the speed of sound inside the pulsar fluid, we obtained the maximum radial sound speed $v_r^2\approx 0.35c^2$ at the center in a better agreement with soft EoSs as predicted by gravitational wave observations. In spite of the capability of the model to lower the speed of sound in comparison to the GR or quadratic gravity with positive $\epsilon$ cases, it is still above the conjectured conformal upper bound on the maximum sound speed $c_s^2=c^2/3$. This result may favor matter-geometry nonminimal coupling scenario to provide better framework for treating that issue as suggested by \cite{ElHanafy:2022kjl,2023arXiv230514953E,Nashed:2023pxd}.
\appendix
\section{The KB model and the induced EoSs}\label{Sec:App_1}

It has been shown that the KB ansatz relates the pressures and the density which effectively induces the EoSs as given by Eqs. \eqref{eq:KB_EoS}. The coefficients in those equations are related to the model parameters as listed as below.
\begin{align}
&c_1=6\,{\frac { \left( -480\,\epsilon_1\,{a_0}^{2}a_2-380\,{a_2}^{3}\epsilon_1-3\,{a_2}^{2}+24\,\epsilon_1\,{a_0 }^{3}+12\,a_2\,a_0+912\,\epsilon_1\,{a_2}^{2}a_0 \right) {c}^{2}}{ \left( 1220\,{a_2}^{3}+384\,{a_0}^{2}a_2-1968\,{a_2}^{2}a_0+72\,
{a_0}^{3} \right) \epsilon_1+15\,{a_2}^{2}}},\label{eq:A1}\\[5pt]
&c_2={\frac {1}{ {\kappa}{R_s}^{2} \left( 4 \left( 305{a_2}^{3}+96{a_0}^{2} a_2-492{a_2}^{2}a_018{a_0}^{3} \right) \epsilon_1+15\,{a_2}^{2} \right)}}\, \bigg\{ \bigg[ 37376\,{a_2}^{4}a_0+1344\,{ a_0}^{4}a_2-38576\,{a_0}^{2}{a_2}^{3}+12000\,{a_2}^{2}{a_0}^{3}\nonumber\\
&-11840\,{a_2}^{5}-1728\,{a_0}^{5} \bigg] {\epsilon_1}^{2}+ \left( -356\,{a_2}^{4}+1408\,{a_2}^{3}a_0+552\,{a_0}^{3}a_2-2040\,{a_0}^{2}{a_2}^{2}+144\,{a_0}^{4} \right) \epsilon_1-6\,{a_2}^{2 } \left( a_0+a_2 \right)  \bigg\},\label{eq:A2}\\[5pt]
&c_3={\frac {\left(-380\,{a_2}^{3}\epsilon_1-3\,{a_2}^{2}+846\,\epsilon_1\,{a_2}^{2}a_0-432\,\epsilon_1\,{a_0}^{2}a_2+9\,a_2\,a_0-3\,{a_0}^{2}+48\,\epsilon_1\,{a_0}^{3}\right) {c}^{2}}{ \left( 1220\,{a_2}^{3}\epsilon_1+15\,{a_2}^{2}+ 384\,\epsilon_1\,{a_0}^{2}a_2-1968\,\epsilon_1\,{a_2}^{2}a_0+72\,\epsilon_1\,{a_0}^{3} \right) ^{2}}},\label{eq:A3}\\[5pt]
&c_4={\frac {1}{ \left( 1220\,{a_2}^{3}\epsilon_1+15\,{a_2}^{2}+384\,\epsilon_1\,{a_0}^{2}a_2-1968\,\epsilon_1\,{a_2}^{2}a_0+72\,\epsilon_1\,{a_0}^{3} \right) {R_s}^{2}{\kappa}}}\bigg\{17800\,{a_2}^{5}{\epsilon_1}^{2}+ \left( -59944\,{\epsilon_1}^{2}a_0+1018\,\epsilon_1 \right) {a_2}^{4}\nonumber\\
&+ \left( 68536\,{\epsilon_1}^{2}{a_0}^{2}-1688\,\epsilon_1\,a_0+3 \right) {a_2}^{3}+ \left( -35160\,{\epsilon_1}^{2}{a_0}^{3}-24\,a_0+174\,\epsilon_1\,{a_0}^{2} \right) {a_2}^{2}+ \left( 10560\,{\epsilon_1}^{2}{a_0}^{4}-276\,\epsilon_1\,{a_0}^{3}+18\,{a_0}^{2} \right) a_2\nonumber\\
&+180\,\epsilon_1\,{a_0}^{4}-2160\,{\epsilon_1}^{2}{a_0}^{5}\bigg\}.\label{eq:A4}
\end{align}
Using the above set of equations, one can find the physical quantities appear in Eq. \eqref{eq:KB_EoS2} in terms of the model parameters, where $v_r^2=c_1$, $\rho_1=\rho_s=-c_2/c_1$, $v_t^2=c_3$ and $\rho_2=-c_4/c_3$.
\section{The density and pressures gradients}\label{Sec:App_2}
Recalling the matter density and the pressures as obtained for the quadratic polynomial $f(R)$ gravity, namely Eqs. \eqref{sol}, we obtain the gradients of these quantities with respect to the radial distance as below
\begin{align}\label{eq:dens_grad}
&\rho'=-\frac{1}{{e^{2\,{\frac {a_2\,{r}^
{2}}{{R_s}^{2}}}}}{r}^{3}{R_s}^{8}{c}^{2}{\kappa}}\, \left\{ 2{R_s}^{4} \left( {R_s}^{4}+3a_2\,
 \left( 3\ell^{2}\epsilon_1-{r}^{2} \right) {R_s}^{2}-2{a_2}^{2}{r}^{4} \right) {e^{{\frac {a_2\,{r}^{2}}{{R_s}^{2}}}
}}-2{R}^{6} \left( {R}^{2}+2\left( {r}^{2}+2{\ell
}^{2}\epsilon_1 \right) a_2 \right) {e^{2\,{\frac {a_2\,{r}^{2}}{{R_s}^{2}}}}}\right.\nonumber\\
&\left.-8{\ell}^{2}\epsilon_1\, \left[ 2{R_s}^{6}a_2+{r}^{2}
 \left( 40{a_2}^{2}-48a_0\,a_2+3{a_0}^{2} \right) {R_s}^{4}-3 \left( 12\,{a_2}^{3}-34a_0\,{a_2}^{2}+13{a_0}^{2}a_2+3{a_0}^{3
} \right) {r}^{4}{R}^{2}\right.\right.\nonumber\\
&\left.\left.-24a_0\,{r}^{6} \left( a_0-a_2
 \right)  \left( 3\,a_2+a_0 \right)  \left( a_0-4a_2 \right)  \right]  \right\},
\end{align}
\begin{align}\label{eq:pr_grad}
  & p'_r=-\frac{1}{{ e^{2\,{\frac {a_2\,{r}^{2}}{{R_s}^{2}}}}}{r}^{3}{R_s}^{8}{ \kappa}}\, \left\{ 2{R_s}^{6} \left[ {R_s}^{2}+ 2\left({r}^{2}+2{\ell} ^{2}\epsilon_1\right) a_2 \right] {e^{2\,{\frac {a_2\,{r}^{2}}{{R_s}^{2}}}}}+2{R_s}^{4} \left[ 12R_s{}^2{\ell}^{2}\epsilon_1\,a_2 -{R_s}^{4}-   \left( a_2+4a_0 \right) {r}^{2} {R_s}^{2}\right.\right.\nonumber\\
&\left.\left.-2a_2\,{r}^{4}a_0 \right] { e^{{\frac {a_2\,{r}^{2}}{{R_s}^{2}}}}}-8{\ell}^{2}\epsilon_1\, \left[4 {R_s}^{6}a_2+ \left( 24a_0\,a_2-11{a_0}^{2}-9\,{a_2}^{2} \right) {r}^{2}{R_s}^{4}-3{r}^{4} a_0\, \left( {a_0}^{2}-9\,a_0\,a_2+6\,{a_2}^ {2} \right) {R_s}^{2}\right.\right.\nonumber\\
&\left.\left.+2{a_0}^{2}{r}^{6} \left( 3\,a_2+a_0 \right)  \left( a_0 -a_2\right)  \right]  \right\},
\end{align}
\begin{align}\label{eq:pt_grad}
&p'_t=\frac{1}{ {e^{2\,{\frac {a_2 \,{r}^{2}}{{R}^{2}}}}}{r}^{3}{R}^{8}{\kappa}} \left\{ 8{R}^{6}a_2\,{e^{2\,{\frac {a_2\,{r} ^{2}}{{R}^{2}}}}}{\ell}^{2}\epsilon_1-2{R}^{2} \left\{  \left[  \left(2a_2-4a_0 \right) {r}^{2}+12{\ell}^{2}\epsilon_1\, \left( a_0-2\,a_2 \right)  \right] {R}^{4}+ \left[  \left({a_2}^{2} -3{a_0}^{2}+a_0\,a_2 \right) {r}^{2}\right.\right.\right.\nonumber\\
&\left.\left.\left.+12{\ell}^{2}\epsilon_1\,a_2\, \left(a_0  -a_2\right)  \right] {r}^{2}{R}^{2}-{r}^{6}a_0\,a_2\, \left(a_0  -a_2\right)  \right\} {e^{{\frac {a_2\,{r}^{2}}{{R}^{2}}}}}-2{\ell}^{2} \left( 4 \left( 3a_0-7a_2 \right) {R}^{6}-12 \left(11 {a_0}^{2}+28{a_2}^{2}-38\,a_0\,a_2 \right) {r}^{2}{R}^{ 4}\right.\right.\nonumber\\
&\left.\left.-12 \left(37a_0\,{a_2}^{2}-25\,{a_0}^{2}a_2-12{a_2}^{3}+12{a_0}^{3} \right) {r} ^{4}{R}^{2}-16{r}^{6}a_0\, \left(a_0 -a_2 \right) \left( 12\,{a_2}^{2}-5\,a_0\,a_2+{a_0}^{2} \right)  \right) \epsilon_1 \right\}.
\end{align}

%

\end{document}